\documentclass[a4,11pt]{article}
\usepackage{graphicx} 
\usepackage{float}
\usepackage{amsmath}
\usepackage{amsfonts}
\usepackage{latexsym}
\usepackage{amssymb}
\makeatletter
 
  \@addtoreset{equation}{section}
 \makeatother

\begin{document}
\title{Making Mean-Variance Hedging Implementable \\ in a Partially Observable Market\\
\it{\Large{-with supplementary contents for stochastic interest rates-}}
~\footnote{
All the contents expressed in this research are solely those of the authors and do not represent any views or 
opinions of any institutions. The authors are not responsible or liable in any manner for any losses and/or damages caused by the use of any contents in this research.
}
%\\ \Large{\it{--Implications of asymmetric CSA and suboptimal strategies--}}
}

\author{Masaaki Fujii\footnote{Graduate School of Economics, The University of Tokyo. e-mail: mfujii@e.u-tokyo.ac.jp},
Akihiko Takahashi\footnote{Graduate School of Economics, The University of Tokyo. e-mail: akihikot@e.u-tokyo.ac.jp}
}
%\begin{center}
\date{
%First version: April 11, 2011\\
First version: June 14, 2013, This version: November 15, 2013%April 6 starts.
%Current version: \today
}
%\end{center}
\maketitle

%%%%%%    TEXT START    %%%%%%

%%%%%%      Macros      %%%%%%
%nakamacro.tex(H120522;0730)
%\documentstyle[11pt]{article}
%\setlength{\textwidth}{10.5in}
%\setlength{\oddsidemargin}{0in}
%\setlength{\topmargin}{-0.52in}
%\setlength{\textheight}{9.0in}
%\setlength{\footskip}{0.7in}

\newtheorem{definition}{Definition}
\newtheorem{assumption}{$[$ A}
\newtheorem{condition}{$[$ C}
\newtheorem{lemma}{Lemma}
\newtheorem{proposition}{Proposition}
\newtheorem{theorem}{Theorem}
\newtheorem{remark}{Remark}
\newtheorem{example}{Example}
\newtheorem{corollary}{Corollary}
%--------------------------------------------------------------------------
%BOLD FACES
\def\n{{\bf n}}
\def\A{{\bf A}}
\def\B{{\bf B}}
\def\C{{\bf C}}
\def\D{{\bf D}}
\def\E{{\bf E}}
\def\F{{\bf F}}
\def\G{{\bf G}}
\def\H{{\bf H}}
\def\I{{\bf I}}
\def\J{{\bf J}}
\def\K{{\bf K}}
\def\L{{\bf L}}
\def\M{{\bf M}}
\def\N{{\bf N}}
\def\O{{\bf O}}
\def\P{{\bf P}}
\def\Q{{\bf Q}}
\def\R{{\bf R}}
\def\S{{\bf S}}
\def\T{{\bf T}}
\def\U{{\bf U}}
\def\V{{\bf V}}
\def\W{{\bf W}}
\def\X{{\bf X}}
\def\Y{{\bf Y}}
\def\Z{{\bf Z}}
\def\cala{{\cal A}}
\def\calb{{\cal B}}
\def\calc{{\cal C}}
\def\cald{{\cal D}}
\def\cale{{\cal E}}
\def\calf{{\cal F}}
\def\calg{{\cal G}}
\def\calh{{\cal H}}
\def\cali{{\cal I}}
\def\calj{{\cal J}}
\def\calk{{\cal K}}
\def\call{{\cal L}}
\def\calm{{\cal M}}
\def\caln{{\cal N}}
\def\calo{{\cal O}}
\def\calp{{\cal P}}
\def\calq{{\cal Q}}
\def\calr{{\cal R}}
\def\cals{{\cal S}}
\def\calt{{\cal T}}
\def\calu{{\cal U}}
\def\calv{{\cal V}}
\def\calw{{\cal W}}
\def\calx{{\cal X}}
\def\caly{{\cal Y}}
\def\calz{{\cal Z}}
%
%YOKUTUKAUMONO
\def\sskip{\hspace{0.5cm}}
\def\simleq{ \raisebox{-.7ex}{\em $\stackrel{{\textstyle <}}{\sim}$} }
\def\leqsim{ \raisebox{-.7ex}{\em $\stackrel{{\textstyle <}}{\sim}$} }
\def\ep{\epsilon}
\def\half{\frac{1}{2}}
\def\iku{\rightarrow}
\def\Iku{\Rightarrow}
\def\ikup{\rightarrow^{p}}
\def\inclusion{\hookrightarrow}
\def\cadlag{c\`adl\`ag\ }
\def\up{\uparrow}
\def\down{\downarrow}
\def\doti{\Leftrightarrow}
\def\douti{\Leftrightarrow}
\def\dochi{\Leftrightarrow}
\def\douchi{\Leftrightarrow}%
%KAIGYOU,ARRAY
\def\yy{\\ && \nonumber \\}
\def\y{\vspace*{3mm}\\}
\def\nn{\nonumber}
\def\be{\begin{equation}}
\def\ee{\end{equation}}
\def\bea{\begin{eqnarray}}
\def\eea{\end{eqnarray}}
\def\beas{\begin{eqnarray*}}
\def\eeas{\end{eqnarray*}}
%
%KONO RONBUN DE TUKAU MONO
\def\hd{\hat{D}}
\def\hv{\hat{V}}
\def\hsd{{\hat{d}}}
\def\hx{\hat{X}}
\def\hsx{\hat{x}}
\def\bsx{\bar{x}}
\def\bsd{{\bar{d}}}
\def\bx{\bar{X}}
\def\ba{\bar{A}}
\def\bb{\bar{B}}
\def\bc{\bar{C}}
\def\bv{\bar{V}}
\def\balpha{\bar{\alpha}}
\def\bbalpha{\bar{\bar{\alpha}}}
\def\combi{\l(\begin{array}{c}\alpha\\ \beta \end{array}\r)}
\def\f{^{(1)}}
\def\s{^{(2)}}
\def\ss{^{(2)*}}
\def\l{\left}
\def\r{\right}
\def\a{\alpha}
\def\b{\beta}
\def\L{\Lambda}
%上に定義されたコマンドは数式モ−ドで用いる。
%--------------------------------------------------

\def\E{{\bf E}}
\def\P{{\bf P}}
\def\Q{{\bf Q}}
\def\R{{\bf R}}

\def\calf{{\cal F}}
\def\calp{{\cal P}}
\def\calq{{\cal Q}}
\def\wtW{\widetilde{W}}
\def\wtB{\widetilde{B}}
\def\wtPsi{\widetilde{\Psi}}
\def\wt{\widetilde}
\def\mbb{\mathbb}
\def\ep{\epsilon}
\def\del{\delta}
\def\part{\partial}
\def\wh{\widehat}
\def\bsigma{\bar{\sigma}}
\def\yy{\\ && \nonumber \\}
\def\y{\vspace*{3mm}\\}
\def\nn{\nonumber}
\def\be{\begin{equation}}
\def\ee{\end{equation}}
\def\bea{\begin{eqnarray}}
\def\eea{\end{eqnarray}}
\def\beas{\begin{eqnarray*}}
\def\eeas{\end{eqnarray*}}
\def\l{\left}
\def\r{\right}
\vspace{10mm}

%%%%%%%%%%%%%%%%%%%%%%%%%%%%%%
\begin{abstract}
%%%%%%%%%%%%%%%%%%%%%%%%%%%%%%
The mean-variance hedging (MVH) problem is studied in a partially observable
market where the drift processes can only be inferred through the observation of asset
or index processes. Although most of the literatures treat the MVH
problem by the duality method, here we study a system consisting of three BSDEs
derived by Mania and Tevzadze (2003) and Mania et.al.(2008) and try to provide
more explicit expressions directly implementable by practitioners. Under the Bayesian
and Kalman-Bucy frameworks, we find that a relevant BSDE can yield a semi-closed
solution via a simple set of ODEs which allow a quick numerical evaluation. This
renders remaining problems equivalent to solving European contingent claims
under a new forward measure, and it is straightforward to obtain a forward looking non-sequential Monte Carlo simulation
scheme. We also give a special example where the hedging
position is available in a semi-closed form.
For more generic setups, we provide explicit expressions of
approximate hedging portfolio by an asymptotic expansion.
These analytic expressions not only allow the hedgers to update the hedging positions
in real time but also make a direct analysis of the terminal distribution of the hedged
portfolio feasible by standard Monte Carlo simulation.
\\
\\
\\
\footnotesize
We have added a brief note on a stochastic short rate and Interest-Rate Futures
to the original version accepted by {\it Quantitative Finance} for publication.

\normalsize
\end{abstract}
\vspace{10mm}
%%%%%%%%%%%%%%%%%%%%%%%%%%%%%%%%%$

\newpage
%%%%%%%%%%%%%%%%%%%%%%%%%%%%%%%%%
\section{Introduction}
%%%%%%%%%%%%%%%%%%%%%%%%%%%%%%%%%
Since the last financial crisis, there are market-wide  efforts for 
standardization of financial products so that they can be traded through
security exchanges or central counterparties.
This is expected to make them more liquid, transparent,
remote from counterparty credit risks, and in particular significantly 
reduces the regulatory cost for financial firms.
For these products, an 
idealistic situation for electronic trading is emerging and 
many financial firms are heavily investing to setup sophisticated e-trading systems
to maintain their profitability for coming years.
At first sight, 
it might appear that it leads the financial market closer to the ideal ``complete" environment.
However, on the other hand, remaining uncleared 
OTC contracts are going to be severely penalized in terms of regulatory cost
so that it gives financial firms a strong incentive to walk away from them.
This inevitably makes a part of security universe less liquid and costlier to trade, 
and can make practitioners reluctant to use them even if they were the most efficient 
hedging instruments before the crisis.
The last crisis also created another complication by
pushing all the practitioners into a new pricing regime for the collateralized contracts.
Growing recognition of the critical importance of 
the choice of collateral and its funding cost makes it impossible to perfectly 
hedge even a very simple cash flow unless one has an easy access to the 
relevant collateral assets or there exist very liquid basis markets.

Considering the above situation, we naturally expect that there
is a growing need of systematic hedging method allowing 
investors to flexibly choose the hedging instruments
based on their own regulatory and accessibility conditions.
Mean-variance hedging (MVH) is a one possible approach to this problem.
MVH has been studied by many authors through duality method and there exist vast literatures on 
the related issues. See, as recent works, Laurent \& Pham (1999)~\cite{Laurent},
Pham (2001)~\cite{Pham} and references therein~\footnote{
See also Pham \& Quenez (2001)~\cite{PhamQuenez} as an application of duality
for the utility maximization in a partially observable market.}.
Although the mathematical understanding of the MVH problem has been greatly progressed
by those adopting the duality,
more practical issues related to the actual implementation of a hedging program
have not attracted much attention so far and there exist only a few special examples 
reported with  explicit expressions. 
In this paper, we try to make a progress in that direction
by studying the system of equations derived by Mania, Tevzadze and 
their co-authors.

In Mania \& Tevzadze (2003)~\cite{MT-MVH}, the authors 
studied a minimizing problem of a  convex cost function
and showed that the optimal value function follows
a backward stochastic partial differential equation (BSPDE).
They have used the flow dynamics of the value function derived from the It\^o-Ventzell formula
combined with a  martingale property of the optimal value function to obtain a 
BSPDE as a sufficient condition for the optimality.
For the MVH problem, they showed that the BSPDE can be decomposed into 
three backward stochastic differential equations (BSDEs).
The technique is extended for a partial information setup by Mania et.al.(2008)~\cite{MTT-MVHP},
for utility maximization by Mania \& Santacroce (2010)~\cite{Mania-S},
and for MVH problem with general semimartingales by
Jeanblanc et.al. (2012)~\cite{Jeanblanc}.

In the following, we consider the MVH problem in a partially observable market 
where the drift processes can only be inferred through the observation 
of stock or any index processes driven by Brownian motions possibly 
with stochastic volatilities. 
Under the Bayesian and Kalman-Bucy frameworks,
we find that a relevant BSDE yields an semi-closed solution 
via a simple set of ODEs allowing a quick numerical evaluation.
This  renders remaining problems equivalent to
solving European contingent claims, and it is 
straightforward to obtain a forward looking 
Monte Carlo simulation scheme using a simple particle method~\cite{FT-particle}.
As far as the optimal hedging positions are concerned, it is also pointed out that
one only needs the standard simulations for the terminal liability and its
Delta sensitivities against the state processes under a certain forward measure.
We also provide explicit expressions for a solvable case and approximate
hedging portfolio for more generic setups by an asymptotic expansion method.
These explicit forms allow the hedgers to update the hedging positions in real time,
and also make the direct analysis of the terminal distribution of 
the hedged portfolio feasible by standard Monte Carlo simulation.
We also provide several numerical examples to demonstrate our procedures.

%%%%%%%%%%%%%%%%%%%%%%%%%%%%%%%%%%%%%%%%%%%%%%%%%%%%%%%%%%%
\section{The Market Setup}
\label{m-setup}
%%%%%%%%%%%%%%%%%%%%%%%%%%%%%%%%%%%%%%%%%%%%%%%%%%%%%%%%%%%
Let $(\Omega, \calf,\bf{P})$ be a complete probability space equipped with 
a filtration $\mbb{F}=\{\calf_t, 0 \leq t \leq T\}$, where $T$ is a fixed time horizon. 
We consider a financial market 
with a risk-less asset, $d$ tradable stocks or indexes $S=\{S_i\}_{1\leq i \leq d}~$, and 
$m:=(n-d)$ non-tradable indexes or otherwise 
state processes relevant for stochastic volatilities $Y=\{Y_j\}_{d+1 \leq j\leq n}$.
For simplicity, we assume that the interest rate is zero in the main body of the paper.
In Section~\ref{sec-IR}, we shall discuss possible extensions with a stochastic interest rate, 
which can be relevant if the hedging target is sensitive to a change of the yield curve.

Using a vector notation of $S$ and $Y$, we write the dynamics of the underlyings as
\bea
\label{S-dynamics}
&&dS_t=\sigma(t,S_t,Y_t)\Bigl(dW_t+\theta_t dt\Bigr)  \\
\label{Y-dynamics}
&&dY_t=\bsigma(t,S_t,Y_t)\Bigl(dW_t+\theta_t dt\Bigr)+
\rho(t,S_t,Y_t)\Bigl(dB_t+\alpha_t dt\Bigr)
\eea
Here, $(W,B)$ are independent $(\bf{P},\calf)$-Brownian motions with 
dimension $d$ and $m$.
$\theta$ and $\alpha$ are $\{\calf_t\}$-adapted market price-of-risk (MPR) processes
for $W$ and $B$.
$\sigma(t,s,y)$, $\bsigma(t,s,y)$ and $\rho(t,s,y)$ 
are assumed to be known smooth functions  
taking values in $\mathbb{R}^{d\times d}$, $\mathbb{R}^{m\times d}$
and $\mathbb{R}^{m\times m}$.
We assume all of them  satisfy the technical conditions to allow
unique strong solutions for $S$ and $Y$.

We denote the available information set for the investor by 
a sub-$\sigma$-field $\calg_t \subset \calf_t$.
We assume that $\mbb{G}=\{\calg_t, 0\leq t \leq T\}$ is the ${\bf{P}}$-augmentation of filtration
generated by the processes of all the stocks $S$ and a subset $\{Y\}^{obs}\subset \{Y_j\}_{d+1\leq j\leq n}$
which are continuously observable assets or indexes but not tradable by the investor by
regulatory or some other reasons.
Although $S$ and $\{Y\}^{obs}$ can be 
observed continuously, we assume that the investor cannot identify their drifts and Brownian 
shocks independently, which is most likely the case in the real financial market.
Thus, neither $\theta$ nor $\alpha$ is  $\{\calg_t\}$-adapted. 
Through the observation 
of quadratic covariation of $S$ and $\{Y\}^{obs}$, 
we can recover the values of $\sigma_t\sigma_t^\top$, $\bsigma^{obs}_t\sigma_t^\top$ and $(\bsigma_t\bsigma_t^\top+
\rho_t\rho_t^\top)^{obs} $ at each time.
We assume the maps $(\sigma,\bsigma, \rho)$ are constructed in such a way that they allow to
fix the values of all the remaining $Y_k \in \{Y\}_{d+1\leq j\leq n}\backslash \{Y\}^{obs}$ uniquely from the values of $\bigl\{S_t,Y_t^{obs}, \sigma_t\sigma_t^\top, \bsigma^{obs}_t\sigma^\top_t, (\bsigma_t\bsigma_t^\top+\rho_t\rho_t^\top)^{obs}\bigr\}$ at every time $t$.
Thus, under the above construction,  the whole elements of $\{Y\}_{d+1\leq j\leq n}$ are in fact $\{\calg_t\}$-adapted.
Let us further assume $\sigma$ and $\rho$ are always nonsingular and thus
\bea
\label{wtW}
\wtW_t&:=&\int_0^t \sigma^{-1}(u,S_u,Y_u)dS_u \nn \\
&=&W_t+\int_0^t \theta_u du  \\
%\eea
%\bea
\wtB_t&:=&\int_0^t \rho^{-1}(u,S_u,Y_u)\Bigl( dY_u-
\bsigma(u,S_u,Y_u)\sigma^{-1}(u,S_u,Y_u)dS_u\Bigr) \nn \\
&=&B_t+\int_0^t \alpha_u du
\label{wtB}
\eea
are actually $\{\calg_t\}$-adapted processes.

%%%%%%%%%%%%%%%%%%%%%%%%%%%%%%%%%%%%%%%%%%%%
\section{Linear Filtering}
%%%%%%%%%%%%%%%%%%%%%%%%%%%%%%%%%%%%%%%%%%%%
From the expressions $(\ref{wtW})$ and $(\ref{wtB})$ and the fact that 
both of $(\wtW,\wtB)$ are observable, we have a linear observation system 
for the MPR processes.  If we further assume that the MPRs 
are either constants or linear Gaussian processes in $(\bf{P},\calf)$, then 
the system becomes a well-known Bayesian or Kalman-Bucy filtering model.
See a textbook written by Bain \& Crisan (2008)~\cite{Crisan} for 
the details of stochastic filtering.

Let us denote
\bea
z_t:=\begin{pmatrix} \theta_t \\ \alpha_t 
\end{pmatrix}, \quad \omega_t=\begin{pmatrix} W_t \\ B_t \end{pmatrix}
\eea
for notational simplicity, and then we put
\bea
\Lambda_t=\exp\Bigl(-\int_0^t z_s^\top d\omega_s -\frac{1}{2}\int_0^t ||z_s||^2 ds\Bigr). 
\eea
For linear filtering models we discuss below, $\Lambda$ is actually shown to be 
 a true $(\bf{P},\calf)$-martingale. We can then define a new measure $\widetilde{P}$ by
\bea
\left.\frac{d\wt{\bf{P}}}{d\bf{P}}\right|_{\calf_t}=\Lambda_t
\eea
then, it is easy to check 
\bea
\wt{\omega}_t:=\begin{pmatrix} \wtW_t \\ \wtB_t \end{pmatrix}
\eea
is a $n$-dimensional $(\wt{\bf{P}},\calf)$-Brownian motion.
By (\ref{S-dynamics}) and (\ref{Y-dynamics}), one can see that $\mbb{G}$ is actually the 
augmented filtration generated by $(\wtW,\wtB)$ (See Ref.~\cite{PhamQuenez} for details.).
$(\wt{\bf{P}},\calf)$-martingale $\wt{\Lambda}_t=1/\Lambda_t$ gives the inverse relation between the measures
\bea
\left. \frac{d {\bf{P}}}{d\wt{\bf{P}}}\right|_{\calf_t}=\wt{\Lambda}_t~.
\eea

We denote the expectation of the MPRs conditional on $\calg_t$ by
\bea
\hat{z}_t:=\begin{pmatrix} \hat{\theta}_t \\ \hat{\alpha}_t \end{pmatrix} 
:=\begin{pmatrix} \mbb{E}[\theta_t|\calg_t] \\ \mbb{E}[\alpha_t|\calg_t]
\end{pmatrix}.
\eea
By Kallianpur-Striebel formula, it is given by
\bea
\hat{z}_t=\frac{\wt{\mbb{E}}[z_t \wt{\Lambda} |\calg_t]}{\wt{\mbb{E}}[\wt{\Lambda}_t|\calg_t]}~.
\label{KSformula}
\eea
where $\wt{\mbb{E}}[~]$ is the expectation under $\wt{\bf{P}}$ measure.
This equation can be explicitly solvable for a Bayesian and also for
a linear Gaussian model.
Note that the processes defined by
\bea
&&N_t=\wtW_t-\int_0^t \hat{\theta}_s ds \\
&&M_t=\wtB_t-\int_0^t \hat{\alpha}_s ds
\eea
are called innovation processes and they are $(\bf{P},\calg)$-Brownian motions.

%%%%%%%%%%%%%%%%%%%%%%%%%%%%%%%%%%%%%%%%%%%%%%%%%%%%
\subsection{A Bayesian model}
%%%%%%%%%%%%%%%%%%%%%%%%%%%%%%%%%%%%%%%%%%%%%%%%%%%%
In this section, we consider a Bayesian model in which the MPR is assumed to be 
$\calf_0$-measurable with a known prior distribution.
The constant vector
\bea
z= \begin{pmatrix} \theta \\ \alpha \end{pmatrix}
\eea
denotes a value of the MPR.
For a concrete calculation, let us assume that $z$ has a prior Gaussian distribution 
with the mean $z_0$ and its  covariance denoted by a positive 
definite symmetric matrix $\Sigma_0$. Let us denote the corresponding density function by $\varsigma(z)$.

In this setup, one has
\bea
\wt{\Lambda}_t=\exp\left(z^\top \wt{\omega}_t-\frac{t}{2} ||z||^2\right)
\eea
and hence
\bea
F(t,\wt{\omega}_t)&:=&\wt{\mbb{E}}[\wt{\Lambda}_t|\calg_t]\nn \\
&=&\int_{\mbb{R}^n}\exp\left(\wt{\omega}_t^\top z-\frac{t}{2}||z||^2\right)\varsigma(z)d^nz~.
\eea
This yields
\bea
\hat{z}_t=\frac{\partial_w F(t,\wt{\omega}_t)}{F(t,\wt{\omega}_t)}.
\eea

For a Gaussian prior distribution, $\hat{z}$ can be evaluated explicitly.
One can show that
\bea
F(t,\wt{\omega}_t)=\frac{1}{(2\pi)^{n/2}|\Sigma_0|^{1/2}}\int
\exp\left(\wt{\omega}_t^\top z-\frac{t}{2}||z||^2-\frac{1}{2}(z-z_0)^\top \Sigma_0^{-1} (z-z_0)
\right)d^nz.
\eea
Using a new positive definite symmetric matrix $\Sigma(t)$ defined by
\bea
\Sigma(t):=[\Sigma_0^{-1}+t\mbb{I}]^{-1}
\label{Bayesian-Sigma}
\eea
and $x:=z-z_0$, one obtains
\bea
F(t,\wt{\omega}_t)=\frac{\exp\Bigl(\wt{\omega}_t^\top z_0-\frac{t}{2}||z_0||^2\Bigr)}{(2\pi)^{n/2}|\Sigma_0|^{1/2}}
\int \exp\Bigl( [\wt{\omega}_t-tz_0]^\top x-\frac{1}{2}x^\top \Sigma(t)^{-1}x\Bigr)d^nx .
\eea
Then, simple calculation gives
\bea
F(t,\wt{\omega}_t)=\sqrt{\frac{|\Sigma(t)|}{|\Sigma_0|}}\exp\left(
-\frac{t}{2}||z_0||^2+\wt{\omega}_t^\top z_0+\frac{1}{2}[\wt{\omega}_t-t z_0]^\top \Sigma(t) [\wt{\omega}_t-tz_0]
\right).
\eea
As a result, the conditional expectation of the MPR is given by
\bea
\hat{z}_t=z_0+\Sigma(t)[\wt{\omega}_t-t z_0]
\eea

Using a simple fact
\bea
\frac{d}{dt}(\Sigma(t)\Sigma(t)^{-1})=0
\eea
one can easily confirm that
\bea
\frac{d}{dt}\Sigma(t)=-\Sigma(t)^2.
\eea
Thus, the dynamics of $\hat{z}$ can be written as
\bea
d\hat{z}_t=-\Sigma(t)\hat{z}_tdt+\Sigma(t)d\wt{\omega}_t
\eea
i.e., 
\bea
d\hat{z}_t=\Sigma(t) dn_t~.
\eea
where we have used a shorthand notation
\be
n_t:=\begin{pmatrix} N_t \\ M_t \end{pmatrix}~.
\ee
Thus, we can see that $\hat{z}$ is a Gaussian martingale process in $(\wt{\bf{P}},\calg)$.

%%%%%%%%%%%%%%%%%%%%%%%%%%%%%%%%%%%%%%%%
\subsection{A Kalman-Bucy model}
%%%%%%%%%%%%%%%%%%%%%%%%%%%%%%%%%%%%%%%%%
In this model, we assume $z_t$ (or, ``signal") follows a linear Gaussian process in $(\bf{P},\calf)$:
\bea
dz_t=[\mu-F z_t]dt+\del dV_t
\label{signal-org}
\eea
where $\mu\in\mathbb{R}^n$ and $\del\in\mbb{R}^{n\times p}, F\in \mathbb{R}^{n\times n}$ are constants.
$V$ denotes $p$-dimensional $(\bf{P},\calf)$-Brownian motions independent of $(W,B)$.
The MPR is assumed to have a prior Gaussian distribution with mean $z_0$ and covariance matrix $\Sigma_0$.

The observation is made through
\bea
d\wt{\omega}_t=z_t dt+d\omega_t~.
\eea

In this case, we have a well-known result that
\bea
d\hat{z}_t=[\mu-F\hat{z}_t]dt+\Sigma(t)dn_t,\quad \hat{z}_0=z_0
\label{KB-dynamics}
\eea
where $\Sigma(t)\in\mbb{R}^{n\times n}$ is a deterministic function given as a solution of 
the following ODE:
\bea
\frac{d\Sigma(t)}{dt}=\del\del^\top -F\Sigma(t)-\Sigma(t)F^\top -\Sigma(t)^2
\label{KB-variance}
\eea
with the initial condition $\Sigma(0)=\Sigma_0$.
We assume that $\Sigma(t)$ is positive definite for all $t\in[0,T]$.
In the remainder of the paper, we provide the detailed calculations only for 
this Kalman-Bucy model. For Bayesian case, one can get the equivalent results by simply putting $\mu=F=0$ 
and using the relevant $\Sigma(t)$ given in (\ref{Bayesian-Sigma}) in 
the corresponding formulas.

Kalman-Bucy scheme still works in the same way with  time-dependent deterministic coefficients 
$(\mu(t),F(t),\delta(t))$. The equations (\ref{KB-dynamics}) and (\ref{KB-variance}) hold true by simply replacing the 
constants with the corresponding time-dependent functions. All the discussions in the paper can be also
extended straightforwardly to this case.
However, throughout  this paper, we treat the constant-coefficients case only.
This is for simplicity and also for the practical difficulty to estimate the time-dependent functions 
using the market data in reality.

\subsubsection*{Remark}
Let us comment on the differences from the related work Pham (2001)~\cite{Pham}, in which 
the author has also worked on MVH problem under Bayesian and Kalman-Bucy frameworks.
In the paper, the author considered the setup where the market observables are
the tradable stocks only, of which the volatility function $\sigma(t,S_t)$ is assumed to be
independent of the non-tradable indexes.
In addition, the hedging target at the maturity $T$ was assumed to be given by 
a function of $S_T$ and $Y$, where $Y$ is $\calf_T$-measurable and 
independent of $S_T$ under the measure ${\bf{P}}$. In our notations, it also means that 
$\hat{\theta}$ is independent of $M$ and that $\hat{\alpha}$ is absent. 
In the current paper, we do not make these simplifying assumptions and deal with practically 
more relevant situations.

%%%%%%%%%%%%%%%%%%%%%%%%%%%%%%%%%%%%%%%%%%%%%%%%%%%%%%
\section{A System of BSDEs for Mean-Variance Hedging}
\label{BSDEs}
%%%%%%%%%%%%%%%%%%%%%%%%%%%%%%%%%%%%%%%%%%%%%%%%%%%%%
Since we are assuming that the interest 
rate is zero, 
the dynamics of wealth with the initial capital $w$ at $s<t$ is given by
\bea
\calw_t^{\pi}(s,w)=w+\int_s^t \pi_u^\top dS_u
\eea
where $\pi\in \Pi$ is a portfolio strategy. Here, 
$\Pi$ denotes a set of $d$-dimensional $\mbb{G}$-predictable processes satisfying 
appropriate integrability conditions.
Our problem is to solve
\bea
V(t,w)={\rm ess}\inf_{\pi\in\Pi}\mathbb{E}\left[
\Bigl(\calw_T^\pi(t,w)-H\Bigr)^2\Bigr|\calg_t\right]~.
\label{V-problem}
\eea
In this paper, we suppose $H$ is some $\calg_T$-measurable (and hence the investor can 
exactly know the terminal liability) square integrable random variable, that is $H \in L^2({\bf P}, \calg_T)$.

Mania \& Tevzadze~\cite{MT-MVH, MTT-MVHP} proved (using more general setup) that
a solution of the above problem is given by
\bea
V(t,w)=w^2 V_2(t)-2 w V_1(t)+V_0(t)
\eea
where $V_2,V_1$ and $V_0$ are the solutions of the following BSDEs:
\bea
&&V_2(t)=1-\int_t^T \frac{||Z_2(s)+V_2(s)\hat{\theta}_s||^2}{V_2(s)}ds-\int_t^T Z_2(s)^\top dN_s
-\int_t^T \Gamma_2(s)^\top dM_s \\
&&V_1(t)=H-\int_t^T \frac{ [Z_2(s)+V_2(s)\hat{\theta}_s]^\top [Z_1(s)+V_1(s)\hat{\theta}_s]}{V_2(s)}ds \nn\\
&&\hspace{65mm}-\int_t^T Z_1(s)^\top dN_s -\int_t^T \Gamma_1(s)^\top dM_s \\
&&V_0(t)=H^2-\int_t^T \frac{||Z_1(s)+V_1(s)\hat{\theta}_s||^2}{V_2(s)}ds-\int_t^T Z_0(s)^\top dN_s
-\int_t^T \Gamma_0(s)^\top dM_s
\eea  
with some positive constant $c$ such that $c<V_2$ under the 
existence of equivalent martingale measures and with some mild conditions.
Here, all the $\{Z_i,\Gamma_i\}$ are $\{\calg_t\}$-adapted processes with 
appropriate dimensionality.

The corresponding optimal wealth process is given by
\bea
&&\calw^{\pi^*}_T(t,w)=w+\int_t^T \frac{[Z_1(s)+V_1(s)\hat{\theta}_s]^\top}{V_2(s)}[dN_s+\hat{\theta}_s ds]\nn \\ 
&&\hspace{30mm}-\int_t^T \calw^{\pi^*}_s(t,w)\frac{[Z_2(s)+V_2(s)\hat{\theta}_s]^\top}{V_2(s)}[dN_s+\hat{\theta}_s ds].
\label{wealth-dynamics}
\eea
Using the relationship
\be
dN_s+\hat{\theta}_s ds= \sigma^{-1}(s,S_s,Y_s)dS_s
\ee
one can easily read off the optimal hedging position from (\ref{wealth-dynamics}) as
\bea
\pi_s^*=(\sigma^{-1})^\top(s,S_s,Y_s)\frac{1}{V_2(s)}\Bigl\{
[Z_1(s)+V_1(s)\hat{\theta}_s]-\calw_s^{\pi^*}[Z_2(s)+V_2(s)\hat{\theta}_s]\Bigr\}~.
\label{hedging-port}
\eea

In our setup with Brownian motions, derivation of the above BSDEs is quite straightforward by 
using It\^o-Ventzell formula and the martingale property of $V(t,\calw_t^{\pi^*})$ for the optimal strategy.
The main ideas are briefly explained in Appendix~\ref{appendix_A}.
The detailed explanation on It\^o-Ventzell formula is available, for example, in the section 3.3 of a textbook~\cite{Kunita}
as a {\it generalized} It\^o formula. It is quite interesting to see  there exists a direct link between the 
BSPDE and the usual HJB equation. 
See discussions given in Mania \& Tevzadze (2008)~\cite{MT-Utility} for this point.

%%%%%%%%%%%%%%%%%%%%%%%%%%%%%%%%%%%%%%%
\section{Solving $V_2$ by ODEs}
%%%%%%%%%%%%%%%%%%%%%%%%%%%%%%%%%%%%%%%
We now try to solve $V_2$ for our Kalman-Bucy filtering model.
Firstly, using the fact that $0<c<V_2$, we transform $V_2,~Z_2$ and $\Gamma_2$ as follows:
\bea
V_L(t)=\log V_2(t), \quad Z_L(t)=Z_2(t)/V_2(t), \quad \Gamma_L(t)=\Gamma_2(t)/V_2(t) ~.
\eea
Simple calculation gives a quadratic growth BSDE
\bea
&&V_L(t)=-\int_t^T \left\{ \frac{1}{2}(||Z_L(s)||^2-||\Gamma_L(s)||^2)+2\hat{\theta}_s^\top 
Z_L(s)+||\hat{\theta}_s||^2\right\}ds \nn \\
&&\qquad-\int_t^T Z_L(s)^\top dN_s-\int_t^T \Gamma_L(s)^\top dM_s~.
\label{qgBSDE}
\eea
The only ingredient of the BSDE is $\hat{z}$ and it has a linear Gaussian form.

Unfortunately, the proof for the existence as well as the
uniqueness of the quadratic growth BSDE (\ref{qgBSDE}) seems unknown at the moment. This is, in particular, 
due to the existence of unbounded MPR  processes in its driver.
In the case of the bounded MPR processes, we can borrow the proof given in Kobylanski (2000)~\cite{Kobylanski},
or by directly treating the BSDE of $(V_2,Z_2)$ as done in Kohlmann \& Tang (2002)~\cite{Kohlmann}.
As we shall see below, we can at least confirm its existence by checking 
the existence of a bounded solution for the Riccati  ordinary differential equation {\it numerically}
for the relevant interval $t\in[0,T]$.

Let us suppose that the solution has the following 
form:
\bea
V_L(t)=\frac{1}{2}\hat{z}_t^\top a^{[2]}(t) \hat{z}_t+a^{[1]}(t)^\top \hat{z}_t+a^{[0]}(t)
\label{hypo}
\eea
where $\{a^{[i]}\}$ are deterministic functions taking values in  $a^{[2]}(t)\in\mbb{R}^{n\times n}$,
$a^{[1]}(t)\in\mbb{R}^n$ and $a^{[0]}(t)\in\mbb{R}$. We can take $a^{[2]}$  as a symmetric form.
Then, simple application of It\^o formula gives
\bea
\begin{pmatrix} Z_L(t) \\ \Gamma_L(t) \end{pmatrix}
=\Sigma(t)[a^{[1]}(t)+a^{[2]}(t)\hat{z}_t]~.
\eea
Substituting this result into (\ref{qgBSDE}), one obtains
\bea
&&dV_L(t)=\Bigl\{\frac{1}{2}a^{[1]}(t)^\top \Xi(t)a^{[1]}(t)+
\Bigl[\Bigl(a^{[2]}(t)\Xi(t)+2{\bf{1}}_{(d,0)}\Sigma(t)\Bigr)a^{[1]}(t)\Bigr]^\top \hat{z}_t \nn \\
&&\quad+\frac{1}{2}\hat{z}_t^\top \Bigl[
2{\bf{1}}_{(d,0)}+a^{[2]}(t)\Xi(t)a^{[2]}(t)+2{\bf{1}}_{(d,0)}\Sigma(t)a^{[2]}(t)+
2a^{[2]}(t)\Sigma(t){\bf{1}}_{(d,0)}\Bigr]\hat{z}_t\Bigr\}dt\nn\\
&&\quad+Z_L(t)^\top dN_t+\Gamma_L(t)^\top dM_t~.
\label{qgBSDE-1}
\eea
Here, we have defined
\bea
\Xi(t):=(\Sigma_d^\top \Sigma_d)(t)-(\Sigma_m^\top \Sigma_m)(t)
\eea
and $\Sigma_d$~($\Sigma_m$) are $d\times n$~($m\times n$) matrices
obtained by restricting to the first $d$  (last $m$) rows of $\Sigma(t)$,
and ${\bf{1}}_{(d,0)}$ is the diagonal matrix which has $1$ for the first $d$ elements
and $0$ for all the others.

On the other hand, the dynamics of $\hat{z}$ in (\ref{KB-dynamics}) and It\^o formula 
yield
\bea
dV_L(t)&=&\Bigl\{ \dot{a}^{[0]}(t)+\mu^\top a^{[1]}(t)+\frac{1}{2}{\rm{tr}}(a^{[2]}(t)\Sigma^2(t))\nn\\
&&+\Bigl[\dot{a}^{[1]}(t)-F^\top a^{[1]}(t)+a^{[2]}(t)\mu\Bigr]^\top \hat{z}_t\nn \\
&&+\frac{1}{2}\hat{z}_t^\top \Bigl[\dot{a}^{[2]}(t)-F^\top a^{[2]}(t)-a^{[2]}(t)F\Bigr]\hat{z}_t\Bigr\}dt\nn \\
&&+Z_L(t)^\top dN_t+\Gamma_L(t)^\top dM_t~.
\eea
Matching the coefficients of $(\hat{z}\hat{z}, \hat{z})$ and a remaining constant term respectively,  
and using the fact that $V_L(T)=0$, one obtains 
the following ODEs~\footnote{Put $\mu=F=0$ and use the corresponding $\Sigma(t)$ for 
our first Bayesian model.}:
\bea
&&\dot{a}^{[2]}(t)=2{\bf{1}}_{(d,0)}+a^{[2]}(t)\Xi(t)a^{[2]}(t)\nn \\
&&\qquad+F^\top a^{[2]}(t)+a^{[2]}(t)F+2\Bigl({\bf{1}}_{(d,0)}\Sigma(t)a^{[2]}(t)
+a^{[2]}(t)\Sigma(t){\bf{1}}_{(d,0)}\Bigr) \\
&&\dot{a}^{[1]}(t)=-a^{[2]}(t)\mu+\Bigl[
F^\top +a^{[2]}(t)\Xi(t)+2{\bf{1}}_{(d,0)}\Sigma(t)\Bigr]a^{[1]}(t) \\
&&\dot{a}^{[0]}(t)=-\mu^\top a^{[1]}(t)-\frac{1}{2}{\rm{tr}}(a^{[2]}(t)\Sigma^2(t))
+\frac{1}{2}a^{[1]}(t)^\top \Xi(t) a^{[1]}(t)
\eea
with terminal conditions $a^{[2]}(T)=a^{[1]}(T)=a^{[0]}(T)=0$.

The ODEs can be solved sequentially in $(a^{[2]}\rightarrow a^{[1]}\rightarrow a^{[0]})$ order.
Due to the quadratic form, the existence of $a^{[2]}$ is not guaranteed 
and it possibly blows up in finite time.
The sufficient conditions for a bounded solution have been intensively studied for this type of
Riccati matrix differential equations. See, for example, Kalman (1960)~\cite{Kalman},
Jacobson (1970)~\cite{Jacobson} and references therein.
In our setting, it requires $ \Xi(t)$ negative semidefinite, which does not hold in general unfortunately.
However, it is still clear that $a^{[2]}(t)$ stays finite in a certain interval around $T$.
As long as $\Sigma(t)$ has a realistic size, the non-blow-up interval seems
wide enough for practical applications in finance. 
In any case, the behavior of $a^{[2]}$ can be 
easily checked numerically.
Once we confirm the boundedness of $a^{[2]}$ (and hence also for $(a^{[1]},a^{[0]})$)
for the relevant interval $t\in[0,T]$, we can see that the (\ref{hypo}) actually
satisfies the BSDE by the standard application of It\^o formula. It then 
 guarantees the existence of the solution for the BSDE (\ref{qgBSDE}).
In fact, this technique for a quadratic BSDE was already discussed in Schroder \& Skiadas (1999)~\cite{SS}
in the application to a recursive utility, but to the best of our knowledge, it is the first time 
as the application to the MVH problem in Mania \& Tevzadze approach.

%%%%%%%%%%%%%%%%%%%%%%%%%%%%%%
\subsubsection*{Remark:}
%%%%%%%%%%%%%%%%%%%%%%%%%%%%%
It is instructive to apply the perturbative solution technique of FBSDEs proposed by
Fujii \& Takahashi (2012)~\cite{FT-analytical} to (\ref{qgBSDE}).
One can confirm that the $V_L$ has a quadratic form of $\hat{z}$ and 
$(Z_L, \Gamma_L)$ have a linear form of $\hat{z}$ at an arbitrary order of the perturbative expansion.
This is actually how we have noticed the existence of a quadratic-form solution.

%%%%%%%%%%%%%%%%%%%%%%%%%%%%%%%%%%%%%%%%%%%%%%%%%%%%%%%%%%%%
\section{$V_1$ as a simple forward expectation of $H$}
%%%%%%%%%%%%%%%%%%%%%%%%%%%%%%%%%%%%%%%%%%%%%%%%%%%%%%%%%%%%
Since the BSDE for $V_1$ is linear
\bea
dV_1(t)=[Z_L(t)+\hat{\theta}_t]^\top [Z_1(t)+V_1(t)\hat{\theta}_t]dt+
Z_1(t)^\top dN_t+\Gamma_1(t)^\top dM_t
\eea
with $V_1(T)=H$, it is clear that we have
\bea
V_1(t)=\mbb{E}^{\cala}\left[ H\exp\Bigl(
-\int_t^T \bigl[ ||\hat{\theta}_s||^2+\hat{\theta}_s^\top Z_L(s)\bigr]ds\Bigr)
\Bigr|\calg_t\right]~.
\eea
Here, the measure $\bf{P}^{\cala}$ is defined by
\bea
\left. \frac{d{\bf{P}}^{\cala}}{d{\bf{P}}}\right|_{\calg_t}=\eta_t
\eea
where
\bea
\eta_t=\exp\Bigl(-\int_0^t [Z_L(s)+\hat{\theta_s}]^\top dN_s-\frac{1}{2}\int_0^t
||Z_L(s)+\hat{\theta}_s||^2ds\Bigr)~.
\eea
By the result of the previous section, $Z_L+\hat{\theta}$ is a linear Gaussian process
and hence the above measure change can be justified, for example, by Lemma 3.9 in \cite{Crisan}.

Now, let us evaluate
\bea
A(t,T):=\mbb{E}^{\cala}\left[ \exp\Bigl(
-\int_t^T \bigl[ ||\hat{\theta}_s||^2+\hat{\theta}_s^\top Z_L(s)\bigr]ds\Bigr)
\Bigr|\calg_t\right]~.
\eea
The argument of $\exp()$ has a quadratic Gaussian form and is given by
\bea
A(t,T)=\mbb{E}^{\cala}\left[\exp\Bigl( -\int_t^T\Bigl\{
\frac{1}{2}\hat{z}_s^\top b^{[2]}(s) \hat{z}_s+b^{[1]}(s)^\top \hat{z}_s\Bigr\}ds\Bigr)
\Bigr|\calg_t\right] 
\label{A-kac}
\eea
where $b^{[2]}(t)\in\mbb{R}^{n\times n}$ and $b^{[1]}(t)\in\mbb{R}^{n}$ are deterministic 
functions defined as
\bea
&&b^{[2]}(t):=2{\bf{1}}_{(d,0)}+{\bf{1}}_{(d,0)}\Sigma(t)a^{[2]}(t)+a^{[2]}(t)\Sigma(t){\bf{1}}_{(d,0)} \\
&&b^{[1]}(t):={\bf{1}}_{(d,0)}\Sigma(t)a^{[1]}(t)~.
\eea
One may notice that the problem is equivalent to the pricing of the zero-coupon bond in 
a quadratic Gaussian short rate model, and we in fact borrow the same technique below.

Let us focus on the Kalman-Bucy model. The result for the Bayesian model
can be  obtained by the simple parameter replacement as before.
In the measure ${\bf{P}}^{\cala}$, the MPR follows
\bea
d\hat{z}_t=[\varphi(t)+\kappa(t)\hat{z}_t]dt+\Sigma(t) dn_t^{\cala}
\eea
where 
\bea
&&\varphi(t):=\mu-(\Sigma_d^\top \Sigma_d)(t)a^{[1]}(t)\nn \\
&&\kappa(t):=-\Bigl[ F+(\Sigma_d^\top \Sigma_d)(t)a^{[2]}(t)+\Sigma(t){\bf{1}}_{(d,0)}\Bigr]
\eea
and $n_t^{\cala}$ is the $({\bf{P}}^\cala,\calg)$-Brownian motion which is related to $n_t$
by Girsanov's theorem as
\bea
n_t^{\cala}=n_t+\int_0^t {\bf{1}}_{(d,0)}\Bigl[\Sigma(s)[a^{[1]}(s)+a^{[2]}(s)\hat{z}_s]+\hat{z}_s\Bigr]ds~.
\eea

Let us suppose $A$ is 
given in the following form~\footnote{The argument $T$ is omitted in $\{c^{[i]}\}$ for 
notational simplicity.}:
\bea
A(t,T)=\exp\left(\frac{1}{2}\hat{z}_t^\top c^{[2]}(t) \hat{z}_t+c^{[1]}(t)^\top \hat{z}_t+c^{[0]}(t)\right)~.
\label{A-sol}
\eea
with deterministic functions $\{c^{[i]}\}$ taking values in
$c^{[2]}(t)\in\mbb{R}^{n\times n}$, $c^{[1]}(t)\in\mbb{R}^n, c^{[0]}(t)\in\mbb{R}$.
From (\ref{A-kac}), one sees that the dynamics of $A$ is given by
\bea
dA(t,T)=A(t,T)\Bigl\{\frac{1}{2}\hat{z}_t^\top b^{[2]}(t) \hat{z}_t+b^{[1]}(t)^\top \hat{z}_t\Bigr\}dt
+(\cdots)dn_t^{\cala}~,
\eea
but from (\ref{A-sol}) and the dynamics of $\hat{z}$ tell us that
\bea
&&dA(t,T)\nn \\
&&=A(t,T)\Bigl\{\frac{1}{2}\hat{z}_t^\top \bigl[\dot{c}^{[2]}(t)+c^{[2]}(t)\kappa(t)+
\kappa(t)^\top c^{[2]}(t)+c^{[2]}(t)\Sigma^2(t)c^{[2]}(t)\bigr]\hat{z}_t\nn \\
&&\quad+\bigl[ \dot{c}^{[1]}(t)+\kappa(t)^\top c^{[1]}(t)+c^{[2]}(t)\varphi(t)+
c^{[2]}(t)\Sigma^2(t)c^{[1]}(t)\bigr]^\top \hat{z}_t\nn\\
&&\quad+\bigl[ \dot{c}^{[0]}(t)+\varphi(t)^\top c^{[1]}(t)+\frac{1}{2}{\rm{tr}}\bigl(c^{[2]}(t)\Sigma^2(t)\bigr)
+\frac{1}{2}c^{[1]}(t)^\top \Sigma^2(t)c^{[1]}(t)\bigr]\Bigr\}dt+(\cdots)dn_t^{\cala}~. \nn \\
\eea

Therefore, one can see that the solution of $A$ is given by the form (\ref{A-sol})
if and only if $\{c^{[i]}\}$ solve the following ODEs:
\bea
&&\dot{c}^{[2]}(t)=b^{[2]}(t)-c^{[2]}(t)\kappa(t)-\kappa(t)^\top c^{[2]}(t)-c^{[2]}(t)\Sigma^2(t)c^{[2]}(t)\\
&&\dot{c}^{[1]}(t)=b^{[1]}(t)-\kappa(t)^\top c^{[1]}(t)-c^{[2]}(t)\varphi(t)-c^{[2]}(t)\Sigma^2(t)c^{[1]}(t)\\
&&\dot{c}^{[0]}(t)=-\varphi(t)^\top c^{[1]}(t)-\frac{1}{2}{\rm{tr}}\bigl(
c^{[2]}(t)\Sigma^2(t)\bigr)-\frac{1}{2}c^{[1]}(t)^\top \Sigma^2(t) c^{[1]}(t)
\eea
with the terminal conditions $c^{[2]}(T)=c^{[1]}(T)=c^{[0]}(T)=0$.
Numerical evaluation can be easily performed in $(c^{[2]}\rightarrow c^{[1]}\rightarrow c^{[0]})$ order.
The solutions of the ODEs have the same problem for their existence due to the quadratic term of $c^{[2]}$ as in the 
case for $a^{[2]}$. For this equation, the non-blow-up conditions
 are satisfied once $b^{[2]}(t)$ is positive  semidefinite (see, \cite{Kalman,Jacobson}.),
which is always possible when $\Sigma(t)$ is sufficiently small. When the condition is not satisfied,
the existence of the solution is dependent on the maturity, in general. 
In the remainder, let us suppose that there exists a finite solution for $(c^{[2]},c^{[1]},c^{[0]})$ in $[0,T]$
for a given parameter set, which can be checked numerically in any case.

If there exists a solution for $\{c^{[i]}\}$, we can define a very useful forward measure ${\bf{P}}^{\cala_T}$ by
\bea
\left. \frac{d{\bf{P}}^{\cala_T}}{d{\bf{P}}^{\cala}}\right|_{\calg_t}=
\frac{A(t,T)}{A(0,T)\exp\Bigl(\int_0^t \bigl[ ||\hat{\theta}_s||^2+\hat{\theta}_s^\top Z_L(s)\bigr]ds\Bigr)}
\eea
under which standard Brownian motion is given by the relation
\bea
n_t^{\cala_T}&=&n_t+\int_0^t {\bf{1}}_{(d,0)}\Bigl\{
\Sigma(s)[a^{[1]}(s)+a^{[2]}(s)\hat{z}_s]+\hat{z}_s\Bigr\}ds \nn \\
&&-\int_0^t \Sigma(s)[c^{[1]}(s)+c^{[2]}(s)\hat{z}_s]ds~
\eea
by Girsanov's theorem.
Using this measure, one can now express $V_1$ in a very simple fashion:
\bea
V_1(t)=A(t,T)\mbb{E}^{\cala_T}\Bigl[H\Bigr|\calg_t\Bigr]~.
\eea
Once we obtained $V_2$ and $V_1$, the optimal capital $w^*$ that achieves the smallest
hedging error at the initial time $t$ is given by 
\bea
w^*=\frac{V_1(t)}{V_2(t)}~.
\eea

%%%%%%%%%%%%%%%%%%%%%%%%%%%%%%%%%%%%%%%%%%%%%%%%%
\section{Monte Carlo Method}
%%%%%%%%%%%%%%%%%%%%%%%%%%%%%%%%%%%%%%%%%%%%%%%%%
In this section, we consider how to evaluate $(V_1,Z_1)$
and $V_0$ by Monte Carlo simulation.  Although $V_0$ is not necessary for the 
specification of the optimal hedging position by Eq.(\ref{hedging-port}), it is needed to obtain the 
optimal value function $V(t,w)$.

For notational simplicity, let us put
\bea
X_t&:=&\begin{pmatrix} S_t \\  Y_t \end{pmatrix}  \\
\gamma(t,X_t)&:=&\begin{pmatrix}
\sigma(t,X_t) & 0 \\
\bsigma(t,X_t) & \rho(t,X_t) \end{pmatrix} ,
\eea
then, the relevant dynamics under $({\bf{P}},\calg)$ can be written as
\bea
dX_t=\gamma(t,X_t)[dn_t+\hat{z}_t dt]~.
\eea
In the forward measure $({\bf{P}}^{\cala_T},\calg)$, it becomes
\bea
dX_t=\gamma(t,X_t)\Bigl\{ dn_t^{\cala_T}+\bigl[\psi(t)+\Psi(t)\hat{z}_t\bigr]dt\Bigr\}
\label{X-dynamics-fwd}
\eea
where $\psi$ and $\Psi$ are deterministic functions given below:
\bea
\psi(t)&:=&\Sigma(t)c^{[1]}(t)-{\bf{1}}_{(d,0)}\Sigma(t)a^{[1]}(t) \\
\Psi(t)&:=&{\bf{1}}_{(0,m)}+\Sigma(t)c^{[2]}(t)-{\bf{1}}_{(d,0)}\Sigma(t)a^{[2]}(t)~.
\eea
Similarly, the dynamics of $\hat{z}$ in $({\bf{P}}^{\cala_T},\calg)$ is given by
\bea
d\hat{z}_t=\bigl[\phi(t)-\Phi(t)\hat{z}_t\bigr]dt+\Sigma(t)dn_t^{\cala_T}
\label{z-dynamics-fwd}
\eea
with deterministic functions $(\phi,\Phi)$:
\bea
\phi(t)&:=&\mu-(\Sigma_d^\top \Sigma_d)(t)a^{[1]}(t)+\Sigma^2(t)c^{[1]}(t) \\
\Phi(t)&:=& F+(\Sigma_d^\top \Sigma_d)(t)a^{[2]}(t)+\Sigma(t){\bf{1}}_{(d,0)}-\Sigma^2(t)c^{[2]}(t)~.
\eea

In the remainder, we consider a situation where the terminal liability $H$ is given
by some function of $X_T$, i.e., 
\bea
H=H(X_T)~.
\eea

%%%%%%%%%%%%%%%%%%%%%%%%%%%%%%%%%
\subsection{Evaluation of $(V_1,Z_1)$}
%%%%%%%%%%%%%%%%%%%%%%%%%%%%%%%%%
Of course, the evaluation of
\bea
V_1(t)=A(t,T)\mbb{E}^{\cala_T}\Bigl[H(X_T)\Bigr|\calg_t\Bigr]
\eea
can be performed by simply running $(X_t,\hat{z}_t)$ under $({\bf{P}}^{\cala_T},\calg)$ 
in standard simulation.

For the evaluation of $Z_1$, we need to introduce the three {\it stochastic flows},
$(\xi_{t,u}, \chi_{t,u}, \wt{\chi}_{t,u})$. 
They are associated with the sensitivity of the values $\hat{z}_u$ and $X_u$ at certain 
future time $u~(>t)$ against the small changes of their initial values at time $t$.
The first one is defined as, for $1\leq i,j \leq n$, 
\bea
(\xi_{t,u})_{i,j}:=\frac{\part \hat{z}_u^j(t,\hat{z})}{\part \hat{z}^i}
\eea
and is actually given as the solution of the following ODE:
\bea
\frac{d\xi_{t,u}}{du}=-\xi_{t,u} \Phi^\top (u),\qquad(\xi_{t,t})_{i,j}=\del_{i,j}~.
\eea
Here, the notation $\hat{z}_u(t,\hat{z})$ emphasizes that $\hat{z}_u$ started from the value $\hat{z}$ at time $t$.

The next two quantities are similarly defined as
\bea
(\chi_{t,u})_{i,j}:=\frac{\part X_u^j(t,x,\hat{z})}{\part x^i},\qquad (\wt{\chi}_{t,u})_{i,j}:=\frac{\part X_u^j(t,x,\hat{z})}{\part \hat{z}^i} ~.
\eea
The three arguments $(t,x,\hat{z})$ indicate that $X$ stated from $x$ at time $t$ but its future value $X_u$
also depends on the value of $\hat{z}$ at time $t$.
One can show that they follow the SDEs
\bea
\label{chi-dynamics}
&&d(\chi_{t,u})_{i,j}=(\chi_{t,u})_{i,k}\part_k \gamma_j(u,X_u)\Bigl\{dn_u^{\cala_T}+[\psi(u)+\Psi(u)\hat{z}_u]du\Bigr\} \\
\label{wtchi-dynamics}
&&d(\wt{\chi}_{t,u})_{i,j}=(\wt{\chi}_{t,u})_{i,k}\part_k\gamma_j(u,X_u)\Bigl\{dn_u^{\cala_T}+[\psi(u)+\Psi(u)\hat{z}_u]du\Bigr\}\nn\\
&&\qquad+(\gamma(u,X_u)\Psi(u))_{j,k}(\xi^\top_{t,u})_{k,i}du
\eea
with initial conditions $(\chi_{t,t})_{i,j}=\del_{i,j}$ and  $\wt{\chi}_{t,t}=0$, respectively.
In the above equations, and also in the reminder of the paper, we will often use the so-called {\it Einstein convention} which assumes
the summation of the duplicated indexes.
For example, (\ref{chi-dynamics}) should be understood to involve $\sum_{k=1}^n$.

Using the above stochastic flows, one obtains 
\bea
&&\begin{pmatrix} Z_1(t) \\ \Gamma_1(t) \end{pmatrix}
= V_1(t) \Bigl(\Sigma(t) [c^{[1]}(t)+c^{[2]}(t)\hat{z}_t]\Bigr) \nn \\
&&+A(t,T)\Bigl\{ 
\mbb{E}^{\cala_T}\bigl[(\chi_{t,T})_{i,j}\part_j H(X_T)|\calg_t\bigr]\gamma_i(t,X_t) 
+\mbb{E}^{\cala_T}\bigl[(\wt{\chi}_{t,T})_{i,j}\part_j H(X_T)|\calg_t\bigr]\Sigma_i(t)\Bigr\}.\nn \\
\label{z1-formula}
\eea
Thus, the simulation of those stochastic flows alongside of the original underlyings $(X,\hat{z})$
provides us the wanted quantity.

%%%%%%%%%%%%%%%%%%%%%%%%%%%%%%%%%%%%%%%%%%%%%%%%%%%%%%%%%%%
\subsubsection*{Remark: Calculation of $Z_1$ from Delta sensitivity}
%%%%%%%%%%%%%%%%%%%%%%%%%%%%%%%%%%%%%%%%%%%%%%%%%%%%%%%%%%%
In the previous formulation, we have introduced the stochastic flows. 
This complication is not avoidable in order to make a {\it one-shot} Monte Carlo 
simulation possible for the evaluation of $V_0$ which will be explained in the next section.
However, if one only needs the hedging position at time $t$ (and if the dimension $n$ is not too large), 
we can take a much simpler approach.
As one can imagine from the definitions of the stochastic flows,
the second line of (\ref{z1-formula}) can also be estimated by the usual ``Delta" 
sensitivity of the terminal liability:
\bea
&&\mbb{E}^{\cala_T}[(\chi_{t,T})_{i,j}\part_j H(X_T)|\calg_t]=\frac{\part}{\part x^i}
\mbb{E}^{\cala_T}[H(X_T)|\calg_t]\nn \\
&&\mbb{E}^{\cala_T}[(\wt{\chi}_{t,T})_{i,j}\part_j H(X_T)|\calg_t]=
\frac{\part}{\part \hat{z}^i}\mbb{E}^{\cala_T}[H(X_T)|\calg_t]~.
\eea
Thus, the required simulations to obtain $(V_1,Z_1)$ are only those for the estimations of 
the terminal liability $H(X_T)$ and its {\it Delta} sensitivities against the underlyings $(X,\hat{z})$
in $({\bf{P}}^{\cala_T},\calg)$ measure.

%%%%%%%%%%%%%%%%%%%%%%%%%%%%%%%%%%%%%%%%%%%%%%%%%%%%%%%%%%%%%%
\subsection{Evaluation of $V_0$}
%%%%%%%%%%%%%%%%%%%%%%%%%%%%%%%%%%%%%%%%%%%%%%%%%%%%%%%%%%%%%%
Let us define, for $(t<s<T)$ and $(1\leq r\leq n)$, 
\bea
&&\calz_s(X_T,\chi_{s,T},\wt{\chi}_{s,T}):=A(s,T)H(X_T)\Bigl\{\Sigma(s)[c^{[1]}(s)+c^{[2]}(s)\hat{z}_s]
+\hat{z}_s\Bigr\}\nn \\
&&+A(s,T)\Bigl\{(\chi_{s,T})_{i,j}\part_jH(X_T)\gamma_i(s,X_s)+
(\wt{\chi}_{s,T})_{i,j}\part_jH(X_T)\Sigma_i(s)~\Bigr\}.
\eea
We also put
\bea
\zeta_1(s):=\begin{pmatrix} Z_1(s) \\ \Gamma_1(s) \end{pmatrix}
\eea
for a lighter notation. Then, it is easy to confirm that
\bea
\zeta_1(s)+V_1(s)\hat{z}_s=\mbb{E}^{\cala_T}\Bigl[ \calz_s(X_T,\chi_{s,T},\wt{\chi}_{s,T}) \Bigr|\calg_s\Bigr]~.
\eea

Note that the Radon-Nikodym derivative between ${\bf{P}}^{\cala_T}$ and ${\bf{P}}$ conditional on $\calg_t$ is given by
\bea
L_t&:=&\left.\frac{d{\bf{P}}^{\cala_T}}{d{\bf{P}}}\right|_{\calg_t}\nn \\
&=&\exp\left(\int_0^t [G(s)+K(s)\hat{z}_s]dn_s-\frac{1}{2}\int_0^t
||G(s)+K(s)\hat{z}_s||^2 ds\right)
\eea
where $G$ and $K$ are the deterministic functions defined as
\bea
G(t)&:=& \Sigma(t)c^{[1]}(t)-{\bf{1}}_{(d,0)}\Sigma(t)a^{[1]}(t)  \\
K(t)&:=& \Sigma(t) c^{[2]}(t)-{\bf{1}}_{(d,0)}\Sigma(t)a^{[2]}(t)-{\bf{1}}_{(d,0)}~.
\eea
Then the inverse relation is given by 
\bea
L_t^{-1}&=&\left.\frac{d{\bf{P}}}{d{\bf{P}}^{\cala_T}}\right|_{\calg_t}\nn \\
&=&\exp\left(-\int_0^t [G(s)+K(s)\hat{z}_s]dn_s^{\cala_T}-\frac{1}{2}\int_0^t
||G(s)+K(s)\hat{z}_s||^2 ds\right)~.
\eea

Since $V_0$ follows a linear BSDE, it is easy to see that $V_0$ satisfies
\bea
V_0(t)=\mbb{E}\left[
H^2(X_T)-\int_t^T e^{-V_L(s)}[\zeta_1(s)+V_1(s)\hat{z}_s]^\top {\bf{1}}_{(d,0)}
[\zeta_1(s)+V_1(s)\hat{z}_s] ds\Bigr|\calg_t\right]~.
\eea
Changing the measure to ${\bf{P}}^{\cala_T}$, one can express it as
\bea
&&V_0(t)=L_t \mbb{E}^{\cala_T}\Bigl[ L_T^{-1} H^2(X_T)-\int_t^T L_s^{-1}e^{-V_L(s)}\mbb{E}^{\cala_T}\bigl[
\calz_s(X_T,\chi_{s,T},\wt{\chi}_{s,T})|\calg_s\bigr]^\top \nn \\
&&\qquad \times{\bf{1}}_{(d,0)}\mbb{E}^{\cala_T}\bigl[\calz_s(X_T,\chi_{s,T},\wt{\chi}_{s,T})|\calg_s\bigr]ds\Bigr|\calg_t\Bigr].~
\label{V0-org}
\eea
Unfortunately, the naive evaluation of the above expression requires sequential Monte Carlo 
simulations and seems numerically too burdensome to be useful in practice.

However, there is a nice way called a {\it particle method}
 to compress convoluted expectations.
The method describes a physical system where multiple copies of particles 
are created at random interaction times following Poisson law. After the creation, 
the particles belonging to a common specie follow the same probability law but are driven by 
independent Brownian motions. This idea was introduced by 
McKean (1975)~\cite{McKean} to solve a certain type of semilinear PDE 
and has been applied to various research areas since then.

For the current problem (\ref{V0-org}), let us introduce a deterministic intensity $\lambda_t$ and 
denote the corresponding random interaction time by $\tau$.
Then, $V_0(t)$ can be represented by
\bea
&&V_0(t)=L_t\mbb{E}^{\cala_T}\Bigl[ L_T^{-1}H^2(X_T)\Bigr|\calg_t\Bigr]\nn \\
&&-{\bf{1}}_{\{\tau>t\}} L_t\mbb{E}^{\cala_T}\Bigl[{\bf{1}}_{\{t<\tau<T\}}L_{\tau}^{-1}
e^{-V_L(\tau)+\int_t^\tau \lambda_u du} \nn \\
&&\times \frac{1}{\lambda_\tau}\Bigl(\calz_\tau(X_T,\chi_{\tau,T},\wt{\chi}_{\tau,T})\Bigr)^{p=1}
{\bf{1}}_{(d,0)} \Bigl(\calz_\tau(X_T,\chi_{\tau,T},\wt{\chi}_{\tau,T})\Bigr)^{p=2} \Bigr|\calg_t\Bigr]~.
\eea
Here, the underlyings (or ``particles") $(X,\hat{z}, \chi,\wt{\chi})$ belong to either the group $(p=1)$ or $(p=2)$, and
they follow the SDEs having the same form (\ref{X-dynamics-fwd}), (\ref{z-dynamics-fwd}),
(\ref{chi-dynamics}) and (\ref{wtchi-dynamics})
respectively, but driven by two independent $n$-dimensional Brownian motions $n^{\cala_T}({p=1})$
and $n^{\cala_T}(p=2)$.
This particle representation 
allows a one-shot non-sequential Monte Carlo simulation.
See Fujii \& Takahashi (2012) \cite{FT-particle} for the details of the particle method as a 
solution technique for BSDEs, and also 
Fujii et.al.(2012) \cite{FST} as a concrete application to the pricing of American options.
\\
\\

As long as there exist solutions for $\{a^{[i]}\}$ and $\{c^{[i]}\}$, the explained procedures
allow us to obtain the solutions for the three BSDEs given in Sec.~\ref{BSDEs} under a quite general setup.
However, it may be tough to update the hedging positions in timely manner in a volatile market, 
and in addition, it seems almost impossible 
to analyze the terminal distribution of the hedged portfolio,
which may be important for financial firms from a risk-management perspective, 
by simulating (\ref{wealth-dynamics}) in the current approach.
In the remainder of the paper, we give an explicitly solvable example and then
an asymptotic expansion method to answer this issue.

%%%%%%%%%%%%%%%%%%%%%%%%%%%%%%%%%%%%%%%%%%
\section{A simple solvable example}
\label{sec-solvable}
%%%%%%%%%%%%%%%%%%%%%%%%%%%%%%%%%%%%%%%%%
In this section, we consider a solvable
case where the terminal liability depends only on a non-tradable index $Y^I\in\{Y\}^{obs}$
\bea
H(X_T)=Y^I_T~.
\eea
Let us suppose that $\gamma^I(t,X_t)=Y^I_t\sigma_y^\top$ where
$\sigma_y\in \mbb{R}^n$ is a $n$-dimensional constant vector.
Then from (\ref{X-dynamics-fwd}),  the index's dynamics under $({\bf{P}}^{\cala_T},\calg)$ can be written as
\bea
dY_s^I=Y_s^I\sigma_y^\top \bigl[\psi(s)+\Psi(s)\hat{z}_s\bigr]ds+Y_s^I \sigma_y^\top dn_s^{\cala_T}~.
\eea
In order to get $V_1$, it is enough to evaluate
\bea
\mbb{E}^{\cala_T}\bigl[Y_T^I|\calg_t\bigr]=Y_t^I\mbb{E}^{\cala_T}\left[
\exp\Bigl(\int_t^T \sigma_y^\top \bigl[\psi(s)+\Psi(s)\hat{z}_s\bigr]ds\Bigr)\Bigr|\calg_t\right]~.
\eea
Since it has an affine structure, one can evaluate the above expectation by the same 
method used for the evaluation of $A(t,T)$.
One can show that
\bea
P(t,T):=\mbb{E}^{\cala_T}\left[
\exp\Bigl(\int_t^T \sigma_y^\top \bigl[\psi(s)+\Psi(s)\hat{z}_s\bigr]ds\Bigr)\Bigr|\calg_t\right]~
\eea
can be written by the deterministic functions ($\beta^{[1]}(t)\in\mbb{R}^n$, $\beta^{[0]}(t)\in\mbb{R}$) and 
$\hat{z}_t$ as
\bea
P(t,T)=\exp\Bigl(\beta^{[1]}(t)^\top \hat{z}_t+\beta^{[0]}(t)\Bigr)
\eea
where $\{\beta^{[i]}\}$ solve the following ODEs:
\bea
&&\dot{\beta}^{[1]}(t)=\Phi(t)^\top \beta^{[1]}(t)-\Psi(t)^\top \sigma_y \\
&&\dot{\beta}^{[0]}(t)=-\phi(t)^\top\beta^{[1]}(t)-\frac{1}{2}\beta^{[1]}(t)^\top \Sigma^2(t)\beta^{[1]}(t)
-\psi(t)^\top \sigma_y
\eea
with terminal conditions $\beta^{[1]}(T)=\beta^{[0]}(T)=0$.

Now, from the above arguments, one obtains
\bea
V_1(t)=Y^I_t A(t,T) P(t,T)~.
\eea
A simple application of It\^o formula gives
\bea
\begin{pmatrix} Z_1(t) \\ \Gamma_1(t) \end{pmatrix}
=V_1(t)\Bigl\{ \sigma_y+\Sigma(t)\bigl[ c^{[1]}(t)+\beta^{[1]}(t)+ c^{[2]}(t)\hat{z}_t\bigr]\Bigr\}~.
\eea
Once we calculate and store all the relevant deterministic functions, it is straightforward to 
evaluate $V_0$ from
\bea
V_0(t)=\mbb{E}\left[(Y_T^I)^2-\int_t^T \frac{||Z_1(s)+V_1(s)\hat{\theta}_s||^2}{V_2(s)}ds
\Bigr|\calg_t\right]
\label{V_0-solvable}
\eea
by standard Monte Carlo simulation.

%%%%%%%%%%%%%%%%%%%%%%%%%%%%%%%%%%%%%%%%%%%%%%%%%%%%%%%%%%%%
\subsection{A numerical test using the solvable example}
\label{sec-solvable-num}
%%%%%%%%%%%%%%%%%%%%%%%%%%%%%%%%%%%%%%%%%%%%%%%%%%%%%%%%%%%%
Let us provide an interesting numerical example which tests 
the consistency of our procedures. 
In this solvable example, we can directly run the optimal 
wealth process $\calw^{\pi^*}_t$ given in $(\ref{wealth-dynamics})$.
Thus, it is possible to compare $V(0,w)=w^2 V_2(0)-2w V_1(0) + V_0(0)$, which is obtained by 
the ODEs and a standard Monte Carlo simulation for (\ref{V_0-solvable}),
with $\mbb{E}[(Y^I_T-\calw_T^{\pi^*})^2]$  directly obtained by running 
the simulation for $Y^I$ and $\calw^{\pi^*}$.

\begin{figure}[htb!]
\begin{center}	
\includegraphics[width=110mm]{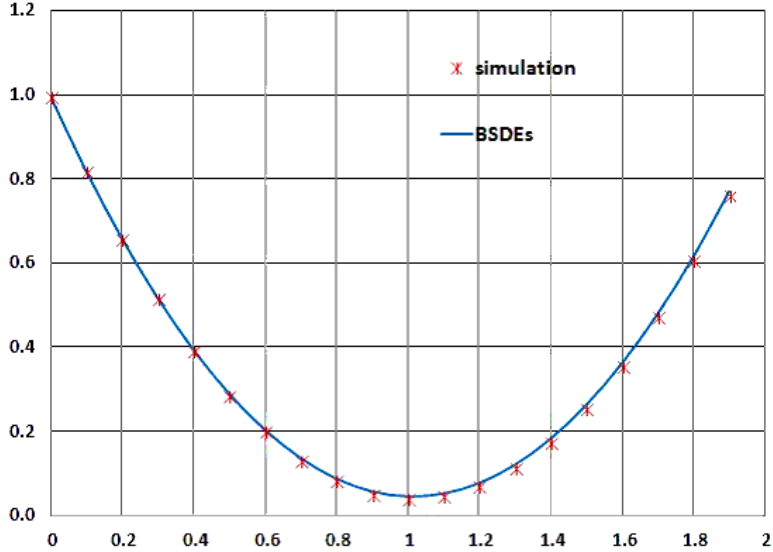}
\end{center}
\caption{Comparison of $V(0,w)=w^2 V_2(0)-2w V_1(0)+V_0(0)$ and direct simulation of $\mbb{E}(Y^I_T-\calw_T^{\pi^*})^2$. 
The solid line is based on the quadratic form of $V(0,w)$ and $\{*\}$ marks are those obtained from the 
direct simulation of the wealth.
The horizontal axis denotes the size of the initial capital $w$. }
\label{analytic-6m}
\end{figure}

\begin{figure}[htb!]
\begin{center}	
\includegraphics[width=110mm]{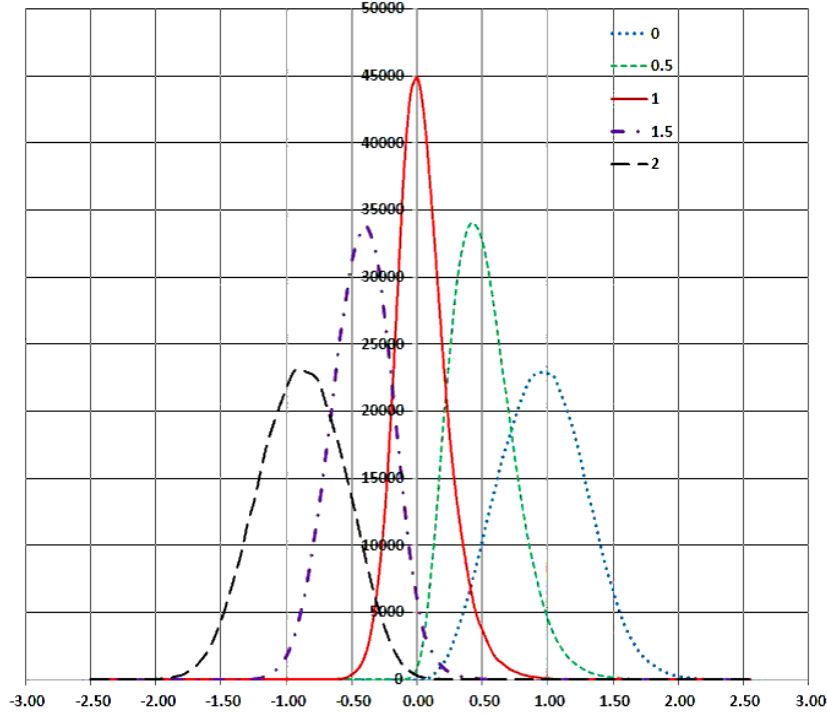}
\end{center}
\caption{The terminal distribution of $(Y^I_T-\calw_T^{\pi^*})$ 
for the five choices of the initial capital $w=\{0,~0.5,~1,~1.5,~2\}$. The graphs are obtained by 
connecting the histograms after sampling $400,000$ paths.}
\label{analytic-6m-dist}
\end{figure}

Let us use the following parameters with $(n=3,~d=2)$:~\footnote{
Here, we put $p=3$ in (\ref{signal-org}). But the choice is free and 
what only matters for the dynamics of $\hat{z}$ is $(\del\del^\top)\in\mbb{R}^{n\times n}$.}
\bea
&&z_0=\begin{pmatrix} 0.3 \\ 0.3 \\ 0.1 \end{pmatrix}, \quad 
\mu=\begin{pmatrix} 0.06 \\ 0.06 \\ 0.02 \end{pmatrix}, \quad 
F=\begin{pmatrix}
0.2 &0.07 &0.05 \\
0.07 &0.2 &0.03 \\
0.05 &0.03& 0.2
\end{pmatrix} \nn \\
&&\del=\begin{pmatrix} 0.3 & 0.15 & -0.1 \\
0.15 & 0.3 & -0.08 \\
-0.03 &-0.07 &0.3 \end{pmatrix}, \quad
\Sigma_0=\begin{pmatrix}
0.2 &0.1 &-0.01 \\
0.1 &0.2 &-0.05  \\
-0.01 &-0.05& 0.2 
\end{pmatrix} ~
\eea
and 
\bea
\sigma_y^\top=(-0.07,-0.12,0.27)~
\eea
with the initial value $Y_0^I=1$.

For $T=0.5$, we have obtained $V_2(0)=0.9263, V_1(0)=0.9399$ by numerically solving ODEs,
and $V_0(0)=0.9974$ after $(100,000+100,000~{\rm{antipathetic}})$ paths with step size $dt=2\times 10^{-3}$.
The standard error for $V_0$ simulation is about $4\times 10^{-4}$.
In Fig.~\ref{analytic-6m}, we have compared the quadratic form of $V(0,w)$
to the results of direct simulation of hedged portfolio with various initial capitals
with the same number of paths and step size for the evaluation of $V_0$.
The standard error for the portfolio simulation is less than $ 4\times 10^{-4}$.
One can see that the prediction of the BSDEs matches very well with the result of the
direct simulation of the hedged portfolio.

One can also study the terminal distribution of the hedged portfolio: $(Y^I_T-\calw_T^{\pi^*})$.
In Fig.~\ref{analytic-6m-dist}, we have plotted the terminal distribution of $(Y^I_T-\calw_T^{\pi^*})$ 
for the five choices of the initial capital $w=\{0,~0.5,~1,~1.5,~2\}$.
The graphs are obtained by connecting the histograms after sampling $400,000$ scenarios with 
the same parameters used to obtain Fig.~\ref{analytic-6m}. One can see distributions of the 
hedged portfolios change consistently with the result of Fig.~\ref{analytic-6m}
and achieves the smallest variance at $w=1$ scenario among the five choices.

%%%%%%%%%%%%%%%%%%%%%%%%%%%%%%%%%%%%%%%%%%%%%%%%%%%%%%%%%%%%%%%%%%%%%%%%%
\section{An asymptotic expansion method}
%%%%%%%%%%%%%%%%%%%%%%%%%%%%%%%%%%%%%%%%%%%%%%%%%%%%%%%%%%%%%%%%%%%%%%%%%
Although it is impossible to obtain a closed-form solution for
\bea
V_1(t)=A(t,T)\mbb{E}^{\cala_T}\bigl[H(X_T) |\calg_t\bigr]
\eea
in general, its evaluation is clearly equivalent to solving a European contingent claim.
Thus, one can borrow various techniques developed for the pricing of financial derivatives 
from the vast existing literatures. Here, we adopt an asymptotic expansion method 
to obtain explicit approximate expressions.
See, for example, \cite{T-org,KT,TTT} and references therein for the details of the method.
In those works, the terminal probability distribution of the underlying process is estimated,
which is then applied to a generic payoff function to price an interested contingent claim.
In this article, however, we adopt a slightly simplified approach in which 
the asymptotic expansion is directly applied to the terminal payoff by assuming $H(x)$
is a smooth function of $x$.  If necessary, we can also apply the original method in \cite{T-org, KT,TTT}
to the current problem, but the resultant formula and required calculation would 
be more involved. We also assume the time-homogeneous volatility 
structure $\gamma(X_t)$ without explicit dependence on $t$ for simplicity.

%%%%%%%%%%%%%%%%%%%%%%%%%%%%%%%%%%%%%%
\subsection{Approximation scheme}
%%%%%%%%%%%%%%%%%%%%%%%%%%%%%%%%%%%%%%
Firstly, let us introduce an auxiliary parameter $\ep$ and $\ep$-dependent processes:
\bea
\label{X-ep}
dX_s^\ep&=&\ep \gamma(X_s^\ep){\bf{1}}_{(0,m)}\hat{z}_s^\ep ds+\ep \gamma(X_s^\ep)dn_s^{\cala_T}\nn \\
&&\qquad+\ep^2 \gamma(X_s^{\ep})[\psi(s)+\wtPsi(s)\hat{z}_s^\ep]ds \\
d\hat{z}_s^\ep&=&\ep\bigl[\phi(s)-\Phi(s)\hat{z}_s^\ep\bigr]ds+\ep \Sigma(s)dn_s^{\cala_T}~,
\label{z-ep}
\eea
where
\bea
\wtPsi(s):=\Sigma(s)c^{[2]}(s)-{\bf{1}}_{(d,0)}\Sigma(s)a^{[2]}(s)~
\label{wtPsi}
\eea
is a deterministic function~\footnote{In theory, there is no need to expand $\hat{z}$ by introducing $\ep$
since it already has a linear dynamics. However, if one treats $\hat{z}$ exactly, 
the calculations associated with $X$ become hugely involved due to the presence of $\hat{z}$ in its drift process
most likely with only a minor improvement of accuracy.}. 
The idea behind this setup is to assume
\bea
\Sigma(s),~\gamma(x),~\mu,~ F
\label{small-params}
\eea
have small enough sizes relative to $\bf{1}$. Then, the auxiliary parameter $\ep$ is introduced to 
count the order of those small quantities appearing in the expansion. 
Since $\Psi$ contains ${\bf{1}}_{(0,m)}$, the 
remaining small term is extracted as $\wtPsi$ in 
(\ref{wtPsi}).

Suppose, we have expanded the $\ep$-dependent process $X^\ep$ as a power series of $\ep$ in the following form:
\bea
X_s^\ep=X_s^{(0)}+\ep X_s^{(1)}+ \ep^2 X_s^{(2)}+\cdots
\label{X-asymptotic}
\eea
where
\bea
X_s^{(k)}:=\left.\frac{1}{k!}\frac{\part^k X_s^\ep}{\part \ep^k} \right|_{\ep=0}~
\eea
Since the term $X^{(k)}$ contains the $k$-th order
products of small quantities in (\ref{small-params}), the higher order terms in (\ref{X-asymptotic})
can be naturally neglected for the approximation purpose. Putting $\ep=1$ at the end of 
calculation provides an approximate valuation for the original process $X$.
In the current work, we will provide the formula for $(V_1,Z_1)$ up to the third order contribution.
The accuracy of approximation is, of course, determined by the size of the quantities given in
(\ref{small-params})~\footnote{More precisely speaking, we need to consider the effect of 
time-integration together.}.
As we can see in the numerical examples provided later in the paper,
the scheme seems to work well with realistic parameters, 
at least for relatively short maturities.

%%%%%%%%%%%%%%%%%%%%%%%%%%%%%%%%%%%%%%%%%%%%%%%%%%%%
\subsection{Asymptotic expansions of the underlying processes}
%%%%%%%%%%%%%%%%%%%%%%%%%%%%%%%%%%%%%%%%%%%%%%%%%%%%
Let us consider the expansion of $(X_s^\ep,\hat{z}_s^\ep)$ for $(s>t)$ under 
the given condition at $t$. 
Obviously, we have
\bea
&&X_s^{(0)}\equiv x \\
&&\hat{z}_s^{(0)}\equiv \hat{z}
\eea
with the conventions that $x:=X_t^\ep$ and $\hat{z}:=\hat{z}_t^\ep$.

Assuming $\gamma(x)$ is smooth enough, one can easily derive
\bea
&&dX_s^{(1)}=\gamma(x){\bf{1}}_{(0,m)}\hat{z}ds+\gamma(x)dn_s^{\cala_T} \\
&&dX_s^{(2)}=\Bigl\{X_s^{i,(1)}\part_i \gamma(x){\bf{1}}_{(0,m)}\hat{z}+
\gamma(x){\bf{1}}_{(0,m)}\hat{z}_s^{(1)}+\gamma(x)[\psi(s)+\wtPsi(s)\hat{z}]\Bigr\}ds\nn \\
&&\hspace{15mm} +X_s^{i,(1)}\part_i\gamma(x)dn_s^{\cala_T} \\
&&dX_s^{(3)}=\Bigl\{
\bigl[ X_s^{i,(2)}\part_i \gamma(x)+\frac{1}{2}X_s^{i,(1)}X_s^{j,(1)}\part_{i,j}\gamma(x)\bigr]
{\bf{1}}_{(0,m)}\hat{z}\\
&&\hspace{15mm}+X_s^{i,(1)}\part_i \gamma(x) {\bf{1}}_{(0,m)}\hat{z}_s^{(1)}+\gamma(x){\bf{1}}_{(0,m)}\hat{z}_s^{(2)} \nn \\
&&\hspace{15mm}+X_s^{i,(1)}\part_i \gamma(x)[\psi(s)+\wtPsi(s)\hat{z}]+\gamma(x)\wtPsi(s)\hat{z}_s^{(1)} 
\Bigr\}ds \nn \\
&&\hspace{15mm}+\bigl[X_s^{i,(2)}\part_i \gamma(x)+\frac{1}{2}\part_{i,j}\gamma(x)\bigr]dn_s^{\cala_T}
\eea
with initial conditions $X_t^{(i)}=0$ for $i\in \{1,2,3\}$. Similarly, for $\hat{z}_s^{(i)}$, one obtains
\bea
&&d\hat{z}_s^{(1)}=[\phi(s)-\Phi(s)\hat{z}]ds+\Sigma(s)dn_s^{\cala_T} \\
&&d\hat{z}_s^{(2)}=-\Phi(s)\hat{z}_s^{(1)}ds \\
&&d\hat{z}_s^{(3)}=-\Phi(s)\hat{z}_s^{(2)}ds
\eea
with $\hat{z}_t^{(i)}=0$ for $i\in\{1,2,3\}$.

%%%%%%%%%%%%%%%%%%%%%%%%%%%%%%%%%%%%%%
\subsubsection{Approximation of $V_1$}
\label{sec-asymp-V1}
%%%%%%%%%%%%%%%%%%%%%%%%%%%%%%%%%%%%%%
Under the assumption that $H(x)$ is smooth enough, one can 
expand it as
\bea
&&\mbb{E}^{\cala_T}\bigl[H(X_T^\ep)|\calg_t\bigr]\nn \\
&&\quad=H(x)+\ep \part_i H(x)\mbb{E}^{\cala_T}\bigl[X_T^{i,(1)}|\calg_t\bigr] \\
&&\quad+\ep^2\Bigl\{
\part_i H(x)\mbb{E}^{\cala_T}\bigl[X_T^{i,(2)}|\calg_t\bigr]+
\frac{1}{2}\part_{i,j}H(x)\mbb{E}^{\cala_T}\bigl[X_T^{i,(1)}X_T^{j,(1)}|\calg_t\bigr]\Bigr\} \\
&&\quad+\ep^3\Bigl\{ \part_i H(x) \mbb{E}^{\cala_T}\bigl[X_T^{i,(3)}|\calg_t\bigr]+
\part_{i,j}H(x)\mbb{E}^{\cala_T}\bigl[X_T^{i,(2)}X_T^{i,(1)}|\calg_t\bigr]\nn \\
&&\qquad\qquad+\frac{1}{6}\part_{i,j,k}H(x)\mbb{E}^{\cala_T}\bigl[
X_T^{i,(1)}X_T^{j,(1)}X_T^{k,(1)}|\calg_t\bigr]\Bigr\}+\calo(\ep^4).
\eea
Since $A(t,T)$ is already available as a solution of the ODEs,
one only needs the expectations of $\{X^{(i)}_T\}$ and their cross products
to obtain an analytic expression of $V_1(t)$.
This is actually calculable  because all the $\{X^{(i)}\}$ have linear dynamics 
thanks to the way we have introduced $\ep$ in (\ref{X-ep}) and (\ref{z-ep}).
Once this is done, 
$Z_1(t)$ can be easily derived by the simple application of It\^o formula.

Let us put
\bea
g(x,\hat{z}):=\gamma(x){\bf{1}}_{(0,m)}\hat{z} {~~~\in\mbb{R}^n}
\eea
and a shorthand notation of a time integration, such as
\bea
&&[f]_t^T :=\int_t^T f(s)ds \nn \\
&&\bigl[ [f]_t^s \bigr]_t^T:=\int_t^T \Bigl(\int_t^s f(u)du \Bigr) ds  \nn \\
&&\hspace{20mm}\cdots
\eea
to lighten the expressions. From the application of It\^o formula,
we can obtain all the necessary expectations as follows:
\bea
&&\mbb{E}^{\cala_T}\bigl[X_T^{(1)}|\calg_t\bigr]=(T-t)g(x,\hat{z}) \nn \\
&&\mbb{E}^{\cala_T}\bigl[X_T^{(2)}|\calg_t\bigr]=
\frac{1}{2}(T-t)^2\part_i g(x,\hat{z})g^i(x,\hat{z})+\gamma(x)\Bigl( [\psi]_t^T+{\bf{1}}_{(0,m)}\bigl[[\phi]_t^s\bigr]_t^T\Bigr)\nn \\
&&\hspace{20mm}+\gamma(x)\Bigl([\wtPsi]_t^T-{\bf{1}}_{(0,m)}\bigl[[\Phi]_t^s\bigr]_t^T\Bigr)\hat{z}  \nn \\
&&\mbb{E}^{\cala_T}\bigl[X_T^{i,(1)}X_T^{j,(1)}|\calg_t\bigr]=(T-t)^2 g^i(x,\hat{z})g^j(x,\hat{z})+(T-t)(\gamma\gamma^\top)_{i,j}(x) \nn
\eea
\bea
&&\mbb{E}^{\cala_T}\bigl[X_T^{(3)}|\calg_t\bigr]=\frac{1}{6}(T-t)^3\Bigl\{\part_i g(x,\hat{z})\part_j g^i(x,\hat{z})g^j(x,\hat{z})
+\part_{i,j}g(x,\hat{z})g^i(x,\hat{z})g^j(x,\hat{z})\Bigr\}\nn \\
&&\hspace{15mm}+\frac{1}{4}(T-t)^2\part_{i,j}g(x,\hat{z})(\gamma\gamma^\top)_{i,j}(x)
+\Bigl(\part_i \gamma(x){\bf{1}}_{(0,m)}\bigl[[\Sigma]_t^s\bigr]_t^T\gamma^\top(x)\Bigr)_i \nn \\
&&\hspace{15mm}+\part_i g(x,\hat{z})\gamma_i(x)\Bigl(\bigl[[\psi]_t^s]_t^T+{\bf{1}}_{(0,m)}\bigl[\bigl[
[\phi]_t^u\bigr]_t^s\bigr]_t^T\Bigr)\nn \\
&&\hspace{15mm}+\part_i g(x,\hat{z})\gamma_i(x)\Bigl(
\bigl[[\wtPsi]_t^s]_t^T-{\bf{1}}_{(0,m)}\bigl[\bigl[[\Phi]_t^u\bigr]_t^s\bigr]_t^T\Bigr)\hat{z}\nn \\
&&\hspace{15mm}+g^i(x,\hat{z})\part_i\gamma(x)\Bigl\{
[(s-t)\psi]_t^T+{\bf{1}}_{(0,m)}\Bigl(\bigl[\bigl[[\phi]_t^u\bigr]_t^s\bigr]_t^T
+\bigl[[(u-t)\phi]_t^s\bigr]_t^T\Bigr)\Bigr\}\nn \\
&&\hspace{15mm}+g^i(x,\hat{z})\part_i\gamma(x)\Bigl\{
[(s-t)\wtPsi]_t^T-{\bf{1}}_{(0,m)}\Bigl(
\bigl[\bigl[[\Phi]_t^u\bigr]_t^s\bigr]_t^T+\bigl[[(u-t)\Phi]_t^s\bigr]_t^T\Bigr)\Bigr\}\hat{z}\nn\\
&&\hspace{15mm}+\gamma(x)\Bigl(\bigl[\wtPsi[\phi]_t^s\bigr]_t^T-{\bf{1}}_{(0,m)}\bigl[\bigl[\Phi[\phi]_t^u\bigr]_t^s\bigr]_t^T
\Bigr)\nn \\
&&\hspace{15mm}+\gamma(x)\Bigl(-\bigl[\wtPsi[\Phi]_t^s\bigr]_t^T+{\bf{1}}_{(0,m)}
\bigl[\bigl[\Phi[\Phi]_t^u\bigr]_t^s\bigr]_t^T\Bigr)\hat{z} \nn
\eea
\bea
&&\mbb{E}^{\cala_T}\bigl[X_T^{i,(2)}X_T^{j,(1)}|\calg_t\bigr]=\frac{1}{2}(T-t)^3 g^k(x,\hat{z})\part_k g^i(x,\hat{z})g^j(x,\hat{z})
+\Bigl(\gamma(x){\bf{1}}_{(0,m)}\bigl[[\Sigma]_t^s\bigr]_t^T\gamma^\top(x)\Bigr)_{i,j}\nn \\
&&\hspace{15mm}+\frac{1}{2}(T-t)^2\Bigl\{g^k(x,\hat{z})((\part_k\gamma)\gamma^\top)_{i,j}(x)
+\part_k g^i(x,\hat{z})(\gamma\gamma^\top)_{k,j}\Bigr\} \nn \\
&&\hspace{15mm}+g^j(x,\hat{z})\gamma_i(x)\Bigl\{
\bigl[[\psi]_t^s\bigr]_t^T+[(s-t)\psi]_t^T+{\bf{1}}_{(0,m)}\Bigl(2\bigl[\bigl[[\phi]_t^u\bigr]_t^s\bigr]_t^T
+\bigl[[(u-t)\phi]_t^s\bigr]_t^T \Bigr)\Bigr\}\nn\\
&&\hspace{15mm}+g^j(x,\hat{z})\gamma_i(x)\Bigl\{
\bigl[[\wtPsi]_t^s\bigr]_t^T+[(s-t)\wtPsi]_t^T-{\bf{1}}_{(0,m)}
\Bigl(2\bigl[\bigl[ [\Phi]_t^u\bigr]_t^s\bigr]_t^T+\bigl[[(u-t)\Phi]_t^s\bigr]_t^T\Bigr)\Bigr\}\hat{z}\nn 
\eea
\bea
&&\mbb{E}^{\cala_T}\bigl[X_T^{i,(1)}X_T^{j,(1)}X_T^{k,(1)}|\calg_t\bigr]\nn \\
&&\hspace{15mm}=(T-t)^2\Bigl\{g^i(x,\hat{z})(\gamma\gamma^\top)_{j,k}(x)+g^j(x,\hat{z})(\gamma\gamma^\top)_{k,i}(x)+g^k(x,\hat{z})(\gamma\gamma^\top)_{i,j}(x)\Bigr\}\nn \\
&&\hspace{15mm}+(T-t)^3g^i(x,\hat{z})g^j(x,\hat{z})g^k(x,\hat{z})\nn \\
\label{upto-3rd-V1}
\eea

Although the expressions are rather lengthy for higher order corrections, there is an 
important feature making our method useful. As one can see from the above result,
the stochastic variable $(x=X_t^\ep,\hat{z}=\hat{z}_t^\ep)$ are separated from all the necessary time integrations.
Thus, one can carry out the required integrations beforehand and store them in the memory,
which then makes possible to use $V_1(t)$ in the simulation with only the usual 
update of underlying state processes $(X_t^\ep, \hat{z}_t^\ep)$. As we shall see next,
this property continues to hold  for $Z_1(t)$.

%%%%%%%%%%%%%%%%%%%%%%%%%%%%%%%%%%%%%%%%%%%%%%%%%%%%%
\subsubsection{Approximation of $(Z_1,\Gamma_1)$}
\label{sec-asymp-Z1}
%%%%%%%%%%%%%%%%%%%%%%%%%%%%%%%%%%%%%%%%%%%%%%%%%%%%%
We now try to expand 
\bea
\zeta_1^\ep(t):=\begin{pmatrix} Z_1^\ep (t) \\ \Gamma_1^\ep(t) \end{pmatrix}
\eea
as
\bea
\zeta_1^\ep(t)=\ep \zeta_1^{(1)}(t)+\ep^2 \zeta_1^{(2)}(t)+\ep^3 \zeta_1^{(3)}(t)+\cdots
\eea
up to the $\ep$-third order corrections.
Since the expansion for 
\bea
V_1^\ep(t)=A(t,T)\mbb{E}^{\cala_T}[H(X^\ep_T)|\calg_t]
\eea
is already obtained, one only needs a simple application of It\^o formula.
Since it increases $\ep$-order by $\bf{1}$, we only need up to the 2nd order corrections
of $V_1^\ep$, and also there is no $0$-th order contribution to $\zeta_1$.

By extracting the coefficients (as {\it row} vector) of the $n$-dimensional 
Brownian motion from the SDEs of the following conditional expectations, 
\bea
&&\overline{X}_{t,T}^{i,(1)}(x,\hat{z}):=\mbb{E}^{\cala_T}[X_T^{i,(1)}|\calg_t] \nn \\
&&\overline{X}_{t,T}^{i,(2)}(x,\hat{z}):=\mbb{E}^{\cala_T}[X_T^{i,(2)}|\calg_t]\nn \\
&&\overline{X}_{t,T}^{(i,j),(1,1)}(x,\hat{z}):=\mbb{E}^{\cala_T}[X_T^{i,(1)}X_T^{j,(1)}|\calg_t]
\eea
one obtains
\bea
&&\bsigma^{i,(1)}_{t,T}(x,\hat{z}):=(T-t)\Bigl\{
\part_j g^i(x,\hat{z})\gamma_j(x) +
\gamma_i(x){\bf{1}}_{(0,m)}\Sigma(t)\Bigr\}  \\
&&\bsigma^{i,(2)}_{t,T}(x,\hat{z}):=\frac{1}{2}(T-t)^2\Bigl[\part_{j,k}g^i(x,\hat{z})g^j(x,\hat{z})+
\part_j g^i(x,\hat{z})\part_k g^j(x,\hat{z})\Bigr]\gamma_k(x)\nn \\
&&\qquad+\frac{1}{2}(T-t)^2\Bigl[g^j(x,\hat{z})\part_j\gamma_i(x)
+\part_jg^i(x,\hat{z})\gamma_j(x)\Bigr]{\bf{1}}_{(0,m)}\Sigma(t) \nn \\
&&\qquad+\part_j \gamma_i(x)\Bigl\{\Bigl([\psi]_t^T+{\bf{1}}_{(0,m)}\bigl[[\phi]_t^s\bigr]_t^T\Bigr)
+\Bigl([\wtPsi]_t^T-{\bf{1}}_{(0,m)}\bigl[[\Phi]_t^s\bigr]_t^T\Bigr)\hat{z}\Bigr\}\gamma_j(x)\nn \\
&&\qquad+\gamma_i(x)\Bigl([\wtPsi]_t^T-{\bf{1}}_{(0,m)}\bigl[[\Phi]_t^s\bigr]_t^T\Bigr)\Sigma(t) \\
&&\bsigma^{(i,j),(1,1)}_{t,T}(x,\hat{z})=(T-t)^2\Bigl[
\part_k g^i(x,\hat{z})g^j(x,\hat{z})+g^i(x,\hat{z})\part_k g^j(x,\hat{z})\Bigr]\gamma_k(x)\nn \\
&&\qquad+(T-t)\Bigl[(\part_k\gamma \gamma^\top)_{i,j}(x)+(\part_k \gamma\gamma^\top)_{j,i}(x)\Bigr]\gamma_k(x)\nn \\
&&\qquad+(T-t)^2\Bigl[g^j(x,\hat{z})\gamma_i(x)+g^i(x,\hat{z})\gamma_j(x)\Bigr]{\bf{1}}_{(0,m)}\Sigma(t)~,
\eea
respectively.
Using this result, one can show that the expansion is finally given by
\bea
&&\zeta_1^{(1)}(t)^\top=A(t,T)H(x)\Bigl[c^{[1]}(t)^\top+\hat{z}^\top c^{[2]}(t)\Bigr]\Sigma(t)+A(t,T)\part_i H(x)\gamma_i(x) \nn \\
%%%%%%%%%%%%%%%%%%%%%%%%%%%%%%%%%%%%%
&&\zeta_1^{(2)}(t)^\top=A(t,T)\Bigl(\part_iH(x)\overline{X}_{t,T}^{i,(1)}(x,\hat{z})\Bigr)\Bigl[c^{[1]}(t)^\top+\hat{z}^\top c^{[2]}(t)\Bigr]\Sigma(t)\nn \\
&&\qquad\qquad+A(t,T)\Bigl[ \part_{i,j}H(x)\overline{X}_{t,T}^{i,(1)}(x,\hat{z})\gamma_j(x)+\part_i H(x)\bsigma^{i,(1)}_{t,T}(x,\hat{z})
\Bigr]\nn \\
%%%%%%%%%%%%%%%%%%%%%%%%%%%%%%%%%%%%%
&&\zeta_1^{(3)}(t)^\top=A(t,T)\bigl[\part_i H(x)\overline{X}^{i,(2)}_{t,T}(x,\hat{z})+
\frac{1}{2}\part_{i,j}H(x)\overline{X}^{(i,j),(1,1)}_{t,T}(x,\hat{z})\bigr]
[c^{[1]}(t)^\top+\hat{z}^\top c^{[2]}(t)]\Sigma(t)\nn\\
&&\qquad\qquad+A(t,T)\Bigl[\part_{i,j}H(x)\overline{X}^{i,(2)}_{t,T}(x,\hat{z})\gamma_j(x)
+\part_i H(x)\bsigma^{i,(2)}_{t,T}(x,\hat{z})\Bigr]\nn \\
&&\qquad\qquad+\frac{A(t,T)}{2}\Bigl[\part_{i,j,k}H(x)\overline{X}^{(i,j),(1,1)}_{t,T}(x,\hat{z})\gamma_k(x)
+\part_{i,j}H(x)\bsigma^{(i,j),(1,1)}_{t,T}(x,\hat{z})\Bigr].\nn\\
\eea

%%%%%%%%%%%%%%%%%%%%%%%%%%%%%%%%%%%%%%%%%%%%%%%%%%%%%%%
\subsection{Numerical Examples}
\label{sec-numerical}
%%%%%%%%%%%%%%%%%%%%%%%%%%%%%%%%%%%%%%%%%%%%%%%%%%%%%%%
As a simple application of the asymptotic expansion, 
let us consider
\bea
H(X_T)=Y^I_T
\eea
as in Sec.~\ref{sec-solvable}, but now 
\bea
\gamma^I(X_t)=(Y^I_t)^\beta \sigma_y^\top~
\eea
for its volatility term. Here, $\beta \in [0,1]$ is some constant, and  $\sigma_y\in\mbb{R}^n$
is a constant vector.
In this case, many cross terms vanish in the asymptotic expansion and 
one obtains rather simple formulas.
The results of the asymptotic expansion for this model are summarized in Appendix~\ref{appendix_B}.

\begin{table}[!htp]
\begin{center}
\begin{tabular}{l|c|c|c|c||c|c|c|c|}																			
\hline
      & $V_1^{(0)}$ & $V_1^{(1)}$ & $V_1^{(2)}$ & $V_1^{(3)}$ & $V_0^{(0)}$ &$V_0^{(1)}$ &$V_0^{(2)}$ & $V_0^{(3)}$ \\ 
\hline
$\beta=0.25$ & 0.87206 & 0.89560 & 0.90216 & 0.90409 & 0.9052 & 1.0095 & 1.0116 & 1.0088 \\
\hline
$\beta=0.5$ & 0.87206 & 0.89560 & 0.90224 & 0.90596 & 0.9106 & 1.0142 & 1.0164 & 1.0160 \\
\hline
\end{tabular}\\
\label{able1}
\caption{
The numerical results for $V_1^{(i)}, V_0^{(i)}$ for $\beta=0.25$ and $\beta=0.5$ models.
$V_1^{(i)}$ is calculated based on the asymptotic expansion {\it including all the 
contribution up to} the $i$-th order. $V_0^{(i)}$ is obtained by running simulation for (\ref{V_0-solvable}) with the corresponding order of approximation for $(V_1,Z_1)$. }
\end{center}
\end{table}

\begin{figure}[!htp]	
\begin{center}
\includegraphics[width=70mm]{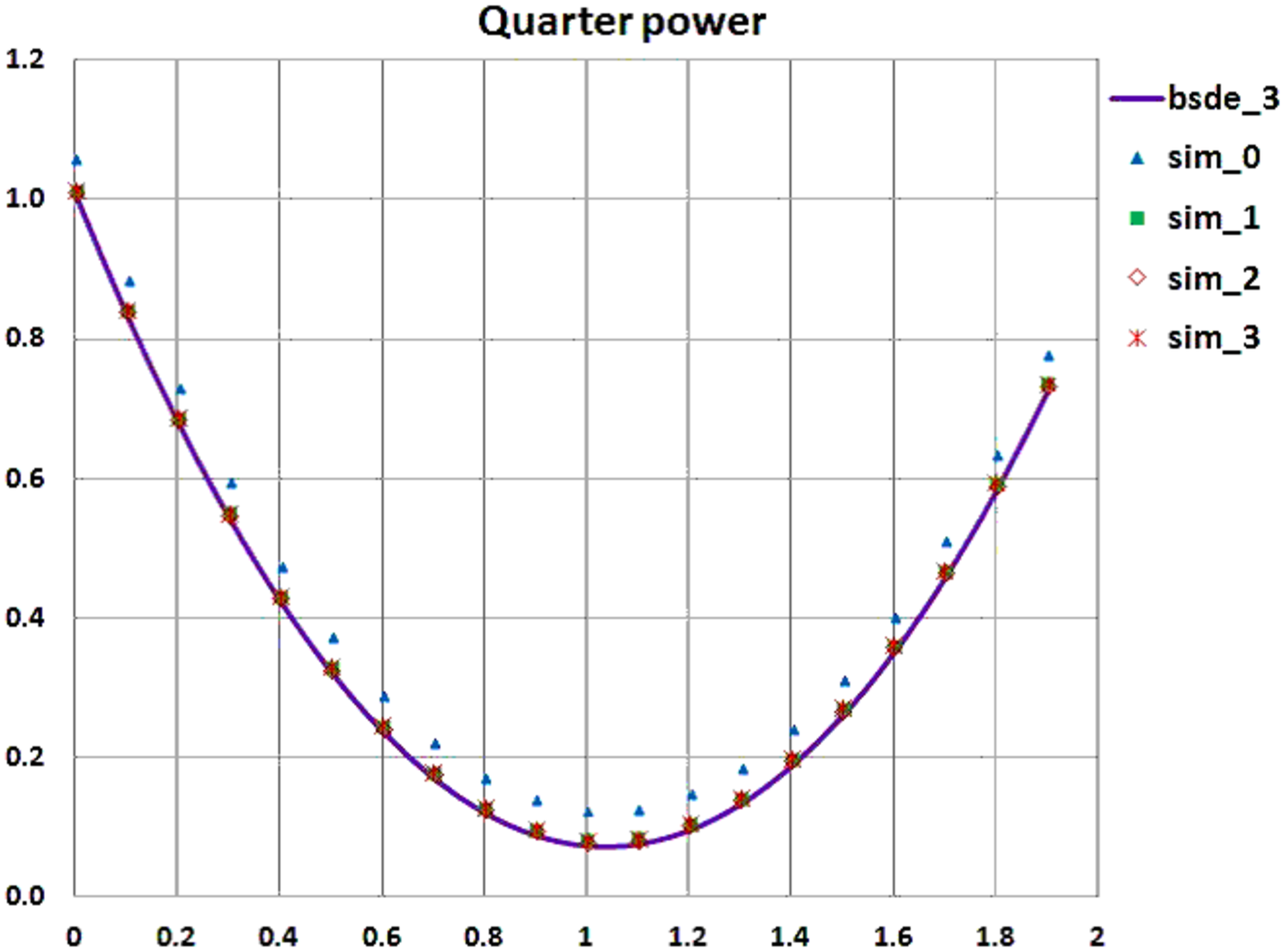}
\includegraphics[width=70mm]{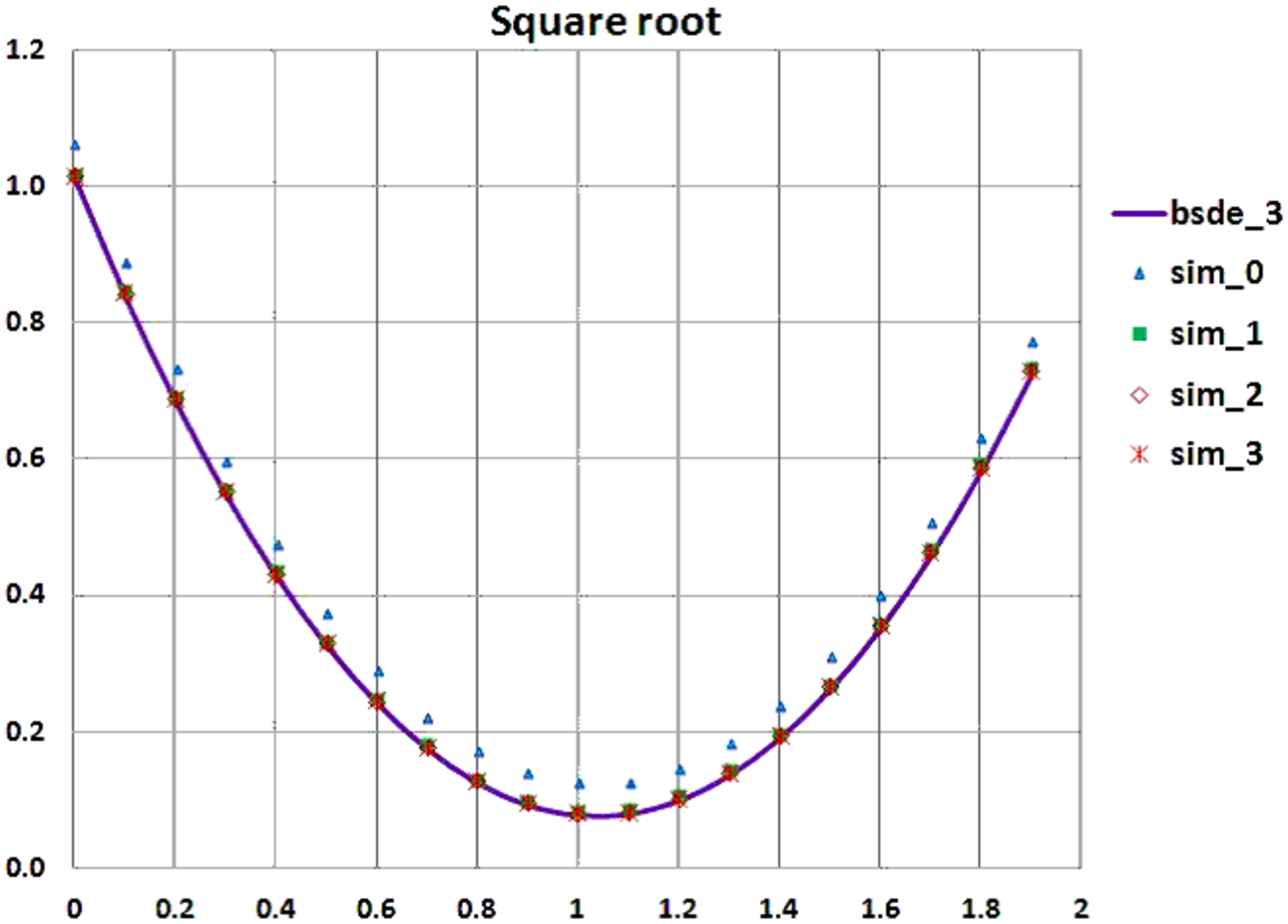}
\end{center}
\caption{Comparison of $V(0,w)\simeq w^2 V_2(0)-2w V_1^{(3)}(0)+V_0^{(3)}(0)$ and direct simulation of $\mbb{E}(Y^I_T-\calw_T^{\pi^*})^2$ with each approximation order of $(V_1,Z_1)$. 
The solid line is based on the quadratic form of $V(0,w)$ and the other symbols are those obtained from the 
direct simulation of wealth with each approximation order.
The horizontal axis denotes the size of the initial capital $w$.}
\label{var-comp-asymptotic}
\end{figure}

We have studied $\beta=0.25$ and $\beta=0.5$ cases for $T=1$yr maturity.
For the remaining parameters $(z_0,~\mu,~F,~\del,~\Sigma_0)$ and also $\sigma_y$ are those
we have used in Sec.~\ref{sec-solvable-num}.
We have also set $Y_0^I=1$ for both of the models.
$V_2(0)$ is independent from the model of $Y^I$ and we have obtained $V_2(0)=0.8721$ by
numerically solving the ODEs.
In Table 1, we have listed the numerical results for $V_1(0)$ and $V_0(0)$.

There, the results of $\{V_1^{(i)}\}$ are based on the asymptotic expansion 
{\it including all the contribution up to} the $i$-th order, and $\{V_0^{(i)}\}$ are
calculated by simulating (\ref{V_0-solvable}) with 
the corresponding order of approximation for $(V_1,Z_1)$.
The number of simulation paths and step size are the same as those used in Sec.~\ref{sec-solvable-num}.
The standard error of $V_0$ simulation is around $7 \times 10^{-4}$ for both of the models.

In Fig.~\ref{var-comp-asymptotic}, we have done the same consistency test as in Sec.~\ref{sec-solvable-num},
where we have compared the quadratic form of $V(0,w)$ and direct simulation of $\mbb{E}(Y^I_T-\calw_T^{\pi^*})^2$.
The solid line corresponds to the prediction of $V(0,w)$ using the 3rd order approximation,
and the other symbols denote the results of direct simulation of $\mbb{E}(Y^I_T-\calw_T^{\pi^*})^2$
using each order of approximation of $(V_1,Z_1)$.
One can confirm the consistency of our approximation and also  
that even the 1st order approximation realizes the most part of 
the hedging benefit of the variance reduction.

\begin{figure}[!hbp]	
\begin{center}
\includegraphics[width=110mm]{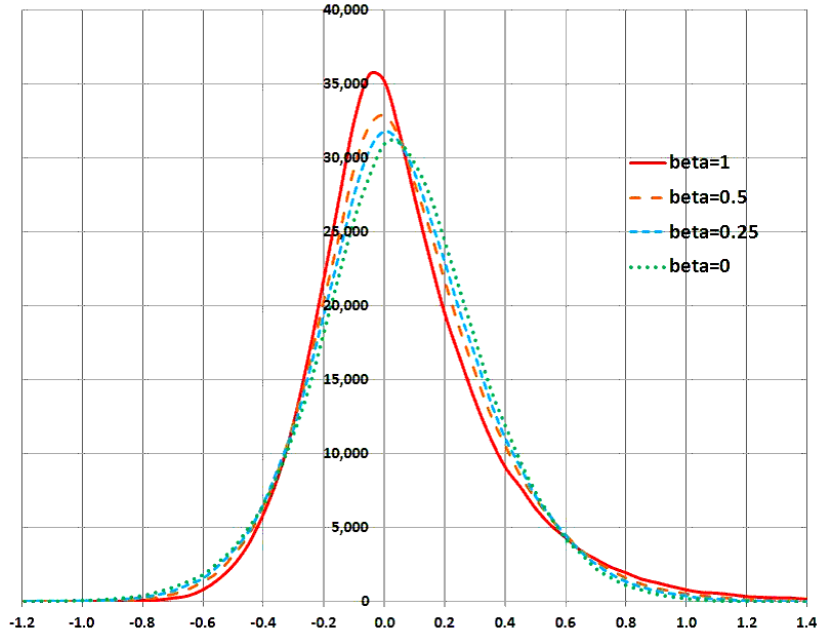}
\end{center}
\caption{The comparison of the terminal distribution $(Y^I_T-\calw_T^{\pi^*})$ with the initial capital $w=1$
using the third order asymptotic expansion. The graphs are obtained by connecting the histograms
after sampling $400,000$ paths. }
\label{dist-comp-asymp}
\end{figure}

It might be surprising that these results are very close between the two choices of $\beta$,
but in fact, this result is naturally expected.
Actually, one can easily confirm that $(V_1^{(0)}(0),~V_1^{(1)}(0))$ should have exactly the same 
value with arbitrary $\beta\in[0,1]$ in the current setup. Furthermore, the common $\sigma_y$ and the initial value of 
$Y^I_0=1$ indicate that every model with $\beta\in[0,1]$ has almost the same variance for relatively short maturities,
which naturally leads to the similar variance for the hedging error.
However, as can be seen from Fig.~\ref{dist-comp-asymp}, there appears a difference in 
the distribution of the hedged portfolio. There, we have compared 
the terminal distribution $(Y^I_T-\calw_T^{\pi^*})$ with the initial capital $w=1$
for four models of $\beta=\{1,~0.5,~0.25,~0\}$.
The graphs of distribution were obtained by connecting the histogram after sampling 400,000
paths. These difference may become important for financial firms from a risk-management perspective.

%%%%%%%%%%%%%%%%%%%%%%%%%%%%%%%%%%%%%%%%%%%%%%%%%%%%%%%%%%%%
\section{An extension to stochastic interest rates}
\label{sec-IR}
%%%%%%%%%%%%%%%%%%%%%%%%%%%%%%%%%%%%%%%%%%%%%%%%%%%%%%%%%%%%
%%%%%%%%%%%%%%%%%
\subsection{Interest-Rate Futures}
%%%%%%%%%%%%%%%%%
Before concluding the paper, we discuss how to handle the situation with 
stochastic interest rates.
Although the effect of discounting is not relevant unless we work on very long term
contracts, the sensitivity of the terminal liability against the change of the forward 
curve of interest rate indexes, such as LIBORs, can be quite 
significant~\footnote{Note that, in the collateralized era after the Lehman default,
LIBORs are not directly related to the discounting rate of the contract.
For cash-collateralized contracts, the corresponding overnight rate is typically
used as the collateral rate and hence as the discounting rate.}.
For this point, we emphasize that our method can be directly applied to 
a futures market, such as Eurodollar futures provided by CME Group in the following way:

Let us introduce the $d$-dimensional ``base" futures prices
$F=(F_i)_{1\leq i \leq d}$ as
\bea
dF(t)=\sigma_F(t,F_t,Y_t)\Bigl(dW_t+\theta_t dt\Bigr)~;\quad F_0={\bf{0}}
\eea
under $(\bf{P},\calf)$. Here, $\sigma_F(t,F_t,Y_t)\in\mbb{R}^{d\times d}$ and $Y$
are assumed to satisfy the same conditions defined in Section~\ref{m-setup}.
Using the base futures prices, we assume that the dynamics of the ``true" futures prices $(L_i)_{i\geq 1}$ 
is given by
\bea
&&L_1(t)=L_1(0)+\int_0^{t\wedge T_1} dF_1(s) \nn \\
&&L_2(t)=L_2(0)+\int_0^{t\wedge T_1} dF_2(s) +\int_{t\wedge T_1}^{t\wedge T_2}dF_1(s) \nn \\
&&\qquad \vdots \nn \\
&&L_d(t)=L_d(0)+\int_0^{t\wedge T_1} dF_d(s)+\cdots +\int_{t\wedge T_{d-1}}^{t\wedge T_d} dF_1(s) \nn \\
&&\qquad \vdots \nn \\
&&L_{d+k}(t)=L_{d+k}(t\wedge T_k)+\int_{t\wedge T_k}^{t\wedge T_{k+1}}dF_d(s)
+\cdots +\int_{t\wedge T_{d+k-1}}^{t\wedge T_{d+k}}dF_1(s)\nn \\
&&\qquad \vdots
\label{true-futures}
\eea
Here, $(T_i)_{i\geq 1}; ~(T_i<T_{i+1})$ denote the set of maturities of the futures contracts.
We consider $d$ as the number of tradable futures at the exchange at any point of time $t$, whose 
underlying's maturities are the $d$ smallest of $\{T_i\}$ larger than $t$.

It is easy to check that, at any point of time $t$, $(F_i(t))_{1\leq i \leq d}$ are
observable, and hence it is possible to adopt the linear filtering scheme in the same way 
as we did for the equity market.
One should notice that the Brownian motions $W$ and the corresponding MPRs $\theta$ are not 
associated with contracts with fixed maturities but rather with the first $d$ rolling contracts.
This setup seems natural since the investor's perception of risk is typically associated with the 
time-to-maturity rather than a specific timing of the maturity.
The dynamics of the wealth by taking a position on the tradable set of $(L_i)$ can be 
equally written by that of the first $d$ rolling set of contracts, $(F_i)_{1\leq i\leq d}$:
\bea
\calw_t^{\pi_F}(s,w)=w+\int_s^t \pi_F^\top (u)dF(u)~.
\eea
Note here that there is no outright cash required to enter and/or  exit a futures contract.
From the expression (\ref{true-futures}), the position on the tradable futures $(L_i)$ 
can be directly read from that on $(F_i)_{1\leq i \leq d}$ at arbitrary point of time.
Neglecting the discounting effect,
it is now clear that one can handle the futures and equity markets completely in 
parallel.

%%%%%%%%%%%%%%%%%%%%%
\subsection{Stochastic Short Rate}
%%%%%%%%%%%%%%%%%%%%%
Lastly, we would like to  mention the fact that introducing a simple stochastic short-rate 
process $(r(t))_{t\geq 0}$ is not difficult as long as it is perfectly observable.
Suppose for example, the money-market account on $r$ is tradable in addition to the 
stocks.
In this case, the wealth-dynamics is given by
\bea
\calw^\pi_t(s,w)=w+\int_s^t \Bigl\{ \pi_u^\top dS_u+\bigl(\calw_u^\pi(s,w)-\pi_u^\top S_u\bigr)r(u)du\Bigr\}~.
\eea
with the modified dynamics of $S$  
\be
dS_t=r(t)S_t dt+\sigma(t,S_t,Y_t)[dW_t+\theta_t dt]
\ee
under $(\bf{P},\calf)$.
Using the above wealth dynamics, the It\^o-Ventzell formula yields
\bea
&&V_2(t)=1-\int_t^T \left\{
\frac{||Z_2(s)+V_2(s)\hat{\theta}_s||^2}{V_2(s)}-2V_2(s)r(s)\right\}ds\nn \\
&&\hspace{35mm}-\int_t^T Z_2(s)^\top dN_s
-\int_t^T \Gamma_2(s)^\top dM_s \\
&&V_1(t)=H-\int_t^T \left\{\frac{[Z_2(s)+V_2(s)\hat{\theta}_s]^\top [Z_1(s)+V_1(s)\hat{\theta}_s]}{V_2(s)}
-V_1(s)r(s)\right\}ds\nn \\
&&\hspace{35mm}-\int_t^T Z_1(s)^\top dN_s-\int_t^T \Gamma_1(s)^\top dM_s
\eea
where the BSDE for $V_0$ is unchanged.
Thus, it only induces additional linear terms $\propto (V_2(s)r(s), V_1(s)r(s))$ to the drivers of $V_2$ and $V_1$, respectively.  

If the short rate process itself is Gaussian, then one can
apply the same technique based on the new state processes $(\hat{\theta}_t,\hat{\alpha}_t, r(t))$.
In the case of a quadratic Gaussian model where
\be
r(t)=A(t)+B(t)^\top X_t+X_t^\top C(t) X_t 
\ee
with some deterministic functions $A(t)\in \mbb{R},B(t)\in \mbb{R}^k,C(t)\in \mbb{R}^{k\times k}$
and $k$-dimensional perfectly observable Gaussian process $(X_t)_{t\geq 0}$, one can use
$(\hat{\theta}_t, \hat{\alpha}_t, X_t)$ instead. 
For these models, adding zero coupon bonds as tradable risky assets is also straightforward.
Although it is unrealistic to assume that the short rate is perfectly observable,
it is still very tightly controlled by the central bank in most of the developed countries.
As long as we work in a relatively short time-horizon, fixing its drift term 
based on the forward guidance provided by the central bank~\footnote{One must assume that the guidance is provided
in the physical measure.}, and allocating all the remaining small daily changes to the 
Brownian motion would be a reasonable approximation.

%%%%%%%%%%%%%%%%%%%%%%%%%%%%%%%%%%%%%%%%%%%%%%%%%
\section{Conclusions}
%%%%%%%%%%%%%%%%%%%%%%%%%%%%%%%%%%%%%%%%%%%%%%%%%
In this article, we have studied the mean-variance hedging (MVH) problem in a partially observable market
by studying a set of three BSDEs derived by Mania \& Tevzadze~\cite{MT-MVH}.
Under the Bayesian and Kalman-Bucy frameworks,
we have found that one of these BSDEs yields a semi-closed solution 
via a simple set of ODEs which allow a quick numerical evaluation.
We have proposed a Monte Carlo scheme using a particle method 
to solve the remaining two BSDEs without nested simulations.
As far as the optimal hedging positions are concerned, it is also pointed out that
one only needs the standard simulations for the terminal liability and its
Delta sensitivities against the state processes under a new  measure 
$({\bf{P}}^{\cala_T},\calg)$. 

We gave a special example where the hedging position is available in a semi-closed form
and presented an interesting consistency test by directly simulating the optimal 
portfolio.
For more general situations, we have provided explicit expressions of the approximate 
hedging portfolio by an asymptotic expansion method and demonstrated the procedures
by several numerical examples.
It would be interesting future works to apply the obtained asymptotic expansion formula to 
more involved situations where the payoff function $H$ is non-linear or dependent on 
both $S$ and $Y$.

Although the simplifying assumptions on the MPR dynamics in $({\bf{P}},\calf)$ are 
very restrictive, generalization to a non-linear dynamics remains as a very challenging 
issue of the non-linear filtering problem with infinite degrees of freedom.
It may be worth considering to use a similar asymptotic expansion technique (see, for example,
Fujii (2013)~\cite{asymp-filtering}.) for this problem.
If the MPR process is perfectly observable, then, in principle, we can take its non-linear effects
into account perturbatively by the method proposed in \cite{FT-analytical}.

\begin{appendix}
%%%%%%%%%%%%%%%%%%%%%%%%%%%%%%%%%%%%%%%%%%%%%%%%%%%%%%%%%%%%%%%%%%
\section{Derivation of BSDEs}
\label{appendix_A}
%%%%%%%%%%%%%%%%%%%%%%%%%%%%%%%%%%%%%%%%%%%%%%%%%%%%%%%%%%%%%%%%%%
In this section, for interested readers, we briefly explain the main ideas of Mania \& Tevzadze leading to the system of BSDEs.
Since $V(t,w)$ defined by (\ref{V-problem}) given the initial capital $w$ at $t$ 
is a $\{\calg_t\}$-adapted semimartingale in general,  
using the {\it ``representation theorem"} (see, Lemma 4.1 of \cite{PhamQuenez}), 
one can decompose it as
\bea
V(t,w)=V(s,w)+\int_s^t a(u,w)du+\int_s^t Z(u,w)^\top dN_u+\int_s^t \Gamma(u,w)^\top dM_u
\label{V-decomp}
\eea
with an appropriate $\{\calg_t\}$-adapted triple $(a,Z,\Gamma)$.
Then, recalling 
\beas
dS_u =\sigma_u[dN_u+\hat{\theta}_u du],
\ \sigma_u\equiv \sigma(u,S_u,Y_u),
\eeas
and assuming appropriate conditions for the use  of It\^o-Ventzell formula~\cite{Kunita},
one obtains 
\bea
&&V(t,\calw_t^\pi)=V(s,w)+\int_s^t a(u,\calw_u^\pi)du+
\int_s^t Z(u,\calw_u^\pi)^\top dN_u+\int_s^t \Gamma(u,\calw_u^\pi)^\top dM_u \nn \\
&&\qquad+\int_s^t V_w(u,\calw_u^\pi)\pi_u^\top \sigma_u[dN_u+\hat{\theta}_u du]+
\int_s^t \pi_u^\top \sigma_u Z_w(u,\calw_u^\pi)du\nn \\
&&\qquad+\int_s^t \frac{1}{2}V_{ww}(u,\calw_u^\pi)\pi_u^\top (\sigma\sigma^\top)(u)\pi_u du
\eea
Here, we have written $\calw_t^\pi(s,w)$ as $\calw_t^\pi$ for simplicity.
It is easy to see $V(t,\calw_t^\pi)$ should be a $(\bf{P},\calg)$-martingale for the optimal strategy $\pi^*$
(and submartingale otherwise). Then, one obtains
\bea
a(s,w)&=&-\inf_{\pi\in\Pi}\left\{\frac{1}{2}V_{ww}(s,w)\Bigl|\Bigl|
\sigma^\top(s)\pi_s+\frac{Z_w(s,w)+V_w(s,w)\hat{\theta}_s}{V_{ww}(s,w)}\Bigr|\Bigr|^2\right\}\nn \\
&&+\frac{||Z_w(s,w)+V_w(s,w)\hat{\theta}_s||^2}{2V_{ww}(s,w)}
\eea
as a drift condition. 

Assuming the $\pi$ which makes the first term zero is admissible and hence corresponding to $\pi^*$, one obtains
\bea
a(s,w)=\frac{||Z_w(s,w)+V_w(s,w)\hat{\theta}_s||^2}{2V_{ww}(s,w)}~.
\eea
Substituting the above result into (\ref{V-decomp}) yields a BSPDE 
\bea
&&V(t,w)=\Bigl|H-w\Bigr|^2-\frac{1}{2}\int_t^T\frac{ ||Z_w(s,w)+V_w(s,w)\hat{\theta}_s||^2}{V_{ww}(s,w)}ds\nn \\
&&\qquad-\int_t^T Z(s,w)^\top dN_s-\int_t^T \Gamma(s,w)^\top dM_s~.
\label{BSPDE}
\eea
The optimal wealth dynamics can also be read as
\bea
\calw_T^{\pi^*}(t,w)=w-\int_t^T \frac{[Z_w(s,\calw_s^{\pi^*})+V_w(s,\calw_s^{\pi^*})\hat{\theta}_s]^\top}{V_{ww}(s,\calw_s^{\pi^*})}
[dN_s+\hat{\theta}_s ds]~.
\eea
Since $\int_t^T( \pi^*(u))^\top dS_u$ is given by the orthogonal projection of $H - w \in L^2(\bf{P})$ 
on the closed subspace of stochastic integrals,
%on the closure in $L^2(\bf{P})$ of the space of stochastic integrals, $\l\{\int_t^T( \pi_u^\top dS_u | \pi \in \Pi\r\}$, 
the optimal strategy $\pi^*$ is linear with
respect to the initial capital $w$.
Thus, one may suppose the following decomposition holds.
(See Theorem 1.4 of \cite{Jeanblanc}  and Theorem 4.1 of \cite{MT-MVH} for the detail.)
\bea
V(t,w)=w^2 V_2(t)-2w V_1(t)+ V_0(t)
\eea
where $\{V_i\}$ do not depend on $w$. 
%(\ref{BSPDE}) 
This decomposition needs to hold for arbitrary $w$. 
Then, inserting back to (\ref{BSPDE}) leads to the 
desired set of BSDEs.  Economic meanings of $V_i$ are explained in \cite{MT-MVH}.

%%%%%%%%%%%%%%%%%%%%%%%%%%%%%%%%%%%%%%%%%%%%%%%%%%%%%%%%%%%%%%%%%%
\section{Asymptotic expansion formulas for the model in Sec.~\ref{sec-numerical}}
\label{appendix_B}
%%%%%%%%%%%%%%%%%%%%%%%%%%%%%%%%%%%%%%%%%%%%%%%%%%%%%%%%%%%%%%%%%%
Firstly, let us put
\bea
&&\overline{Y}_{t,T}^{I,(1)}(y,\hat{z}):=\mbb{E}^{\cala_T}[Y^{I,(1)}_T|\calg_t]\nn \\
&&\overline{Y}_{t,T}^{I,(2)}(y,\hat{z}):=\mbb{E}^{\cala_T}[Y^{I,(2)}_T|\calg_t]\nn \\
&&\overline{Y}_{t,T}^{I,(3)}(y,\hat{z}):=\mbb{E}^{\cala_T}[Y^{I,(3)}_T|\calg_t]~,
\eea
with the convention that
\bea
y:=Y^{I,\ep}_t~.
\eea

From the results in Sec.~\ref{sec-asymp-V1} and \ref{sec-asymp-Z1},
one obtains
\bea
&&\overline{Y}_{t,T}^{I,(1)}(y,\hat{z})=(T-t)y^\beta (\sigma_y^\top {\bf{1}}_{(0,m)}\hat{z}) \nn\\
&&\overline{Y}_{t,T}^{I,(2)}(y,\hat{z})=\frac{1}{2}(T-t)^2\beta y^{2\beta-1}(\sigma_y^\top {\bf{1}}_{(0,m)}\hat{z})^2+y^\beta \sigma_y^\top \Bigl( [\psi]_t^T+{\bf{1}}_{(0,m)}\bigl[[\phi]_t^s\bigr]_t^T\Bigr)\nn \\
&&\qquad +y^\beta \sigma_y^\top \Bigl([\wtPsi]_t^T-{\bf{1}}_{(0,m)}\bigl[[\Phi]_t^s\bigr]_t^T\Bigr)\hat{z}\nn\\
&&\overline{Y}^{I,(3)}_{t,T}(y,\hat{z})\nn \\
&&\quad=\frac{1}{6}(T-t)^3(2\beta^2-\beta)y^{3\beta-2}(\sigma_y^\top 
{\bf{1}}_{(0,m)}\hat{z})^3+\frac{1}{4}(T-t)^2(\beta^2-\beta)y^{3\beta-2}(\sigma_y^\top{\bf{1}}_{(0,m)}\hat{z})||\sigma_y||^2\nn\\
&&\quad+\beta y^{2\beta-1}(\sigma_y^\top{\bf{1}}_{(0,m)}\hat{z})\sigma_y^\top
\Bigl\{\bigl[[\psi]_t^s\bigr]_t^T+[(s-t)\psi]_t^T
+{\bf{1}}_{(0,m)}\Bigl(2\bigl[\bigl[[\phi]_t^u\bigr]_t^s\bigr]_t^T+\bigl[[(u-t)\phi]_t^s\bigr]_t^T\Bigr)\Bigr\}\nn \\
&&\quad+\beta y^{2\beta-1}(\sigma_y^\top {\bf{1}}_{(0,m)}\hat{z})\sigma_y^\top
\Bigl\{\bigl[[\wtPsi]_t^s\bigr]_t^T+[(s-t)\wtPsi]_t^T
-{\bf{1}}_{(0,m)}\Bigl(2\bigl[\bigl[[\Phi]_t^u\bigr]_t^s\bigr]_t^T+\bigl[[(u-t)\Phi]_t^s\bigr]_t^T\Bigr)\Bigr\}\hat{z}\nn \\
&&\quad+y^\beta\sigma_y^\top \Bigl(
\bigl[\wtPsi[\phi]_t^s\bigr]_t^T-{\bf{1}}_{(0,m)}\bigl[\bigl[\Phi[\phi]_t^u\bigr]_t^s\bigr]_t^T\Bigr)
+y^\beta \sigma_y^\top \Bigl(-\bigl[\wtPsi[\Phi]_t^s\bigr]_t^T+{\bf{1}}_{(0,m)}\bigl[\bigl[\Phi[\Phi]_t^u\bigr]_t^s
\bigr]_t^T\Bigr)\hat{z}\nn \\
&&\quad+\beta y^{2\beta-1}\bigl( \sigma_y^\top {\bf{1}}_{(0,m)}\bigl[[\Sigma]_t^s\bigr]_t^T\sigma_y\bigr)~.
\eea
Using the above results, one can show $V_1^\ep(t)$ can be expanded as
\bea
V_1^\ep(t)=A(t,T)\Bigl\{y+\ep \overline{Y}^{I,(1)}_{t,T}(y,\hat{z})
+\ep^2 \overline{Y}^{I,(2)}_{t,T}(y,\hat{z})+\ep^3
\overline{Y}^{I,(3)}_{t,T}(y,\hat{z})+o(\ep^3)\Bigr\}~.
\eea
It is also straightforward to obtain
\bea
&&\zeta_1^{(1)}(t)^\top=A(t,T)\Bigl\{ y[c^{[1]}(t)^\top+\hat{z}^\top c^{[2]}(t)]\Sigma(t)+y^\beta \sigma_y^\top\Bigr\}\nn\\
&&\zeta_1^{(2)}(t)^\top=A(t,T)\Bigl\{\overline{Y}^{I,(1)}_{t,T}(y,\hat{z})
[c^{[1]}(t)^\top+\hat{z}^\top c^{[2]}(t)]\Sigma(t)
+\bsigma^{I,(1)}_{t,T}(y,\hat{z})\Bigr\}\nn \\
&&\zeta_1^{(3)}(t)^\top=A(t,T)\Bigl\{ \overline{Y}^{I,(2)}_{t,T}(y,\hat{z})[c^{[1]}(t)^\top+\hat{z}^\top c^{[2]}(t)]\Sigma(t)
+\bsigma^{I,(2)}_{t,T}(y,\hat{z})\Bigr\}~,
\eea
with the definitions of
\bea
&&\bsigma^{I,(1)}_{t,T}(y,\hat{z}):=(T-t)\Bigl\{
\beta y^{2\beta-1}(\sigma_y^\top {\bf{1}}_{(0,m)}\hat{z})\sigma_y^\top
+y^\beta (\sigma_y^\top{\bf{1}}_{(0,m)}\Sigma(t))\Bigr\} \nn \\
&&\bsigma^{I,(2)}_{t,T}(y,\hat{z})=\frac{1}{2}(T-t)^2(2\beta^2-\beta)y^{3\beta-2}
(\sigma_y^\top{\bf{1}}_{(0,m)}\hat{z})^2\sigma_y^\top \nn \\
&&\qquad+(T-t)^2\beta y^{2\beta-1}(\sigma_y^\top {\bf{1}}_{(0,m)}\hat{z})(\sigma_y^\top 
{\bf{1}}_{(0,m)}\Sigma(t))\nn \\
&&\qquad+\beta y^{2\beta-1}\sigma_y^\top \Bigl[
\Bigl([\psi]_t^T+{\bf{1}}_{(0,m)}\bigl[[\phi]_t^s\bigr]_t^T\Bigr)
+\Bigl([\wtPsi]_t^T -{\bf{1}}_{(0,m)}\bigl[[\Phi]_t^s\bigr]_t^T\Bigr)\hat{z}\Bigr]\sigma_y^\top\nn \\
&&\qquad+y^\beta \sigma_y^\top \Bigl( [\wtPsi]_t^T-{\bf{1}}_{(0,m)}\bigl[[\Phi]_t^s\bigr]_t^T
\Bigr)\Sigma(t).\nn\\
\eea

\end{appendix}

%%%%%%%%%%%%%%%%%%%%%%%%%%%%%%%%%%%%%%%%%%%%%%%%%%%%%%%%%%%%%%%%%%%%%
\section*{Acknowledgement}
%%%%%%%%%%%%%%%%%%%%%%%%%%%%%%%%%%%%%%%%%%%%%%%%%%%%%%%%%%%%%%%%%%%%%
This research is partially supported by Center for Advanced Research in Finance (CARF).
%%%%%%%%%%%%%%%%%%%%%%%%%%%%%%%%%%%%%%%%%%%%%%%%%%%%%%%%%%%%%%%%%%%%%%%%%%%%%

%%%%%%%%%%%%%%%%%%%%%%%%%%%%%%%%%%%%%%%%%%%%%%%%%%%%%%%%%%%%%%%%%%%%

\end{document}